\documentclass[structabstract]{aa}
\pdfoutput=1
\usepackage{graphicx}
\usepackage{natbib}
\newcommand{\p}[1]{{\color{blue}{#1}}}

\newcommand{\fig}[1]{Figure~\ref{#1}}

\newcommand{\jetone}{{Jet 1}}
\newcommand{\jettwo}{{Jet 2}}
\begin{document}
\title{Twisting solar coronal jet launched at the boundary of an active region}
\author{B. Schmieder\inst{1}
\and
 Y. Guo\inst{2}
\and
 F. Moreno-Insertis \inst{3,4}
\and
 G. Aulanier \inst{1}
\and
 L. Yelles Chaouche \inst{3,4}
\and
 N. Nishizuka \inst{5,6}
\and
 L. K. Harra \inst{6}
\and
 J. K. Thalmann \inst{7}
\and
 S. Vargas Dominguez \inst{8}
\and
 Y. Liu \inst{9}
}

\institute
{LESIA, Observatoire de Paris, CNRS, UPMC, Universit\'e
  Paris Diderot, 5 place Jules Janssen, 92190 Meudon, France 
\and School of Astronomy and Space Science, Nanjing
University, Nanjing 210093, China 
\and Instituto de Astrofisica de Canarias, Via Lactea, s/n,
38205 La Laguna (Tenerife), Spain 
\and Dept.~of Astrophysics, Universidad de La Laguna, 38200 La Laguna
(Tenerife), Spain
\and National Astronomical Observatory of Japan, Mitaka, Tokyo 181-8588, Japan
\and UCL-Mullard Space Science Laboratory, Holmbury
St. Mary, Dorking, Surrey, RH5 6NT, UK 
\and Max-Plank-Institut f\"ur Sonnensystemforschung,
Max-Planck-Str. 2, 37191 Katlenburg-Lindau, Germany 
\and Departamento de F\'isica, Universidad de Los Andes,
A.A. 4976, Bogot\'a, Colombia 
\and W. W. Hansen Experimental Physics Laboratory, Stanford
University, Stanford, CA 94305, USA 
}

\date{Received date / Accepted date }
\titlerunning{Twisting solar Coronal Jet Launched by Bald Patch Magnetic Reconnection}
\authorrunning{Schmieder et al.}

\abstract
{}
{A broad jet was observed in a weak magnetic  field area at the edge
  of active region NOAA 11106 that also produced other
  nearby recurring and narrow jets.  The peculiar shape and
  magnetic environment of the broad jet raised the question
  of whether it was created by the same physical processes
  of previously studied jets with reconnection occurring
  high in the corona.  }
{We carried out a multi-wavelength analysis using the EUV
  images from the Atmospheric Imaging Assembly (AIA) and
  magnetic fields from the Helioseismic and Magnetic Imager
  (HMI) both  on-board the \textit{SDO} satellite, which we
  coupled to a high-resolution, nonlinear force-free field
  extrapolation.  Local correlation tracking was used to
  identify the photospheric motions that triggered the jet,
  and time-slices were extracted along and across the jet to
  unveil its complex nature. A topological analysis of the
  extrapolated field was performed and was related to the
  observed features.  }
{The jet consisted of many different threads that expanded in around 10
  minutes to about $100$ Mm in length, with the bright features in later
  threads moving faster than in the early ones, reaching a maximum speed of
  about 200 km~s$^{-1}$.  Time-slice analysis revealed a striped pattern of
  dark and bright strands propagating along the jet, along with apparent
  damped oscillations across the jet. This is suggestive of a (un)twisting
  motion in the jet, possibly an Alfv\'en wave. Bald patches in field lines, low-altitude flux ropes, diverging flow patterns,  and a null point
  were identified at the basis of the jet.  }
{Unlike classical $\lambda$ or Eiffel-tower shaped jets that
  appear to be caused by reconnection in current sheets
  containing null points, reconnection in regions containing
  bald patches seems to be crucial in triggering the present
  jet. There is no observational evidence that the flux
  ropes detected in the topological analysis were actually
  being ejected themselves, as occurs in the violent phase
  of blowout jets; instead, the jet itself may have gained
  the twist of the flux rope(s) through reconnection. This event may
  represent a class of jets different from the classical quiescent or blowout
  jets, but to reach that conclusion, more observational and
    theoretical work is necessary.}


\maketitle

\section{Introduction}

Solar coronal jets were first identified in X-rays using data
from the Soft X-ray Telescope (SXT; \citealt{1991Tsuneta}) on-board
the Japanese mission {\it Yohkoh}. A statistical study determined that
their typical length is $\approx$ few times $10^4$-- $4 \times
10^5$~km, the width $5 \times 10^3$--$10^5$~km, the apparent
average velocity around 200~km~s$^{-1}$, and the lifetime from a
few minutes to ten hours \citep{1996Shimojo}.  Another analysis
of 100 X-ray jets determined the magnetic configurations at the
base of the jets: 8\% of the jets occur in a single magnetic
polarity, 12\% in bipoles, 24\% in mixed polarities, and 48\% in
a satellite polarity \citep{1998Shimojo}.  The thermal parameters
of jet plasma were intensively studied by \citet{2000Shimojo},
who found the correlation between the temperatures of the jets
and the sizes of the flares at the footpoints.  Surges are
observed in chromospheric spectral lines with a brightening at
their base. Ultraviolet (UV) and X-ray jets may be associated
with surges. Some attempts have been made at correlating X-ray jets
and surges \citep[e.g.,][]{1995Schmieder,1996Canfield}.  It was
shown that cold and hot plasma exist along  different magnetic field
lines.  The energy and temperature of the jet plasma were found
to be related to the intensity of micro-flares.

More recently, some space-borne missions, such as \textit{TRACE},
\textit{STEREO}, and \textit{Hinode}, have provided high resolution
observations of jets in a multi-wavelength range with many
components. Type II spicules with high velocities were
detected in the Ca II line with the Solar Optical Telescope (SOT)
onboard \textit{Hinode} and were found to be associated with
velocity jets observed with the Extreme-ultraviolet Imaging
Spectrometer (EIS) on the same satellite. \citet{2009McIntosh}
proposed that these combined jets  provide enough energy to
heat the coronal plasma. \citet{2011Ugarte-Urra} analyzed jets at
the periphery of active regions in cool (Si VII) and hot (Fe XII)
lines observed with Hinode/EIS and concluded that the hot and
cold plasmas are not  directly related to one another. More
powerful jets have been observed with data from the X-Ray
Telescope (XRT) onboard \textit{Hinode} with high kinetic
energies \citep{2007Cirtain,2007Savcheva}. Jets and surges are
often associated with micro-flares and X-ray bright points
\citep{1995Schmieder,1996Mandrini}, with anemone-shaped
configurations for coronal loops
\citep{1992Shibata265,2012Zhang}. Some of the observed jets
display helical features that  may be interpreted as associated
with untwisting field line motions
\citep[e.g., ][]{2008Patsourakos, 2011Shen, 2012Chen}.

The observed features as well as previous theoretical
considerations \citep{1977Heyvaerts} indicated that jets are
caused by magnetic reconnection, which led to the first dynamical
model, in two spatial dimensions, in which the reconnection was
driven by flux emergence (\citealt{1995Yokoyama,1996Yokoyama,
  2008Nishizuka}). At present,  X-ray or EUV jets are
being modeled in three dimensions, with the jets resulting either
from the interaction of an emerging magnetic bipole with the
preexisting coronal field \citep{2008Moreno-Insertis,
  2012Archontis, 2013Moreno-Insertis} or from the effect of
horizontal photospheric motions onto a simple null-point
configuration in the atmosphere \citep{2009Pariat, 2010Pariat},
with, in all those cases, an open magnetic field in the
background corona. In both types of models, the jet is a
consequence of the reconnection taking place in a current sheet
at the interface between the perturbed and preexisting magnetic
domains. The current sheet contains at least one null point and,
as shown by \citet{2013Moreno-Insertis}, may have a complex
topology, including plasmoids.  Formation of null points in the
corona has also been shown by \citet{2009Torok} using a
zero-$\beta$ MHD simulation: the interface between an emerging
twisted flux tube and a pre-existing arcade contains a null-point
with associated reconnection that leads to a torsional Alfv\'en
wave and the development of a sheared loop system.

Of particular interest is the recent identification of a subclass of X-ray
and EUV jets called {\it blowout jets} (\citealt{2010Moore, 2010Sterling};
see also \citealt{2011Liu, 2011A&A...526A..19M}).  This class shows a first
phase with a regular quiescent jet, followed by a violent flux-rope eruption
at its base, whereby cool material at chromospheric or transition region
temperatures is ejected, perhaps a cool rope as in a mini-CME eruption. An
interpretation of this phenomenon has been offered  recently
\citep{2013Moreno-Insertis, 2013ApJ...769L..21A} using the
 3D flux-emergence models cited above. In the
first of those references, the base of a quiescent jet becomes unstable and
causes a collection of several flux rope ejections through different
mechanisms (tether-cutting reconnection, kink instability) occurring in
different parts of the emerged-field domain below the jet, which links that
model with previous MHD simulations of the tether-cutting instability
\citep{2004ApJ...610..588M, 2012Archontis} or of the kink instability
\citep{2003A&A...406.1043T} in coronal flux ropes.

Of the various fundamental questions that are still open
concerning X-ray / EUV jets in the solar corona, this paper
is related with the following ones:

\begin{itemize}
\item
are there structural differences between jets occurring within
coronal holes, and jets that take place at their boundary, or at
the boundary of active regions?

\item 
are blowout jets a distinct subclass of X-ray / EUV jets, or do
most quiescent jets develop a violent phase with flux rope
eruptions toward the end of their lives? Vice versa: is there a
class of jets consisting only of a violent flux rope ejection,
that is,  a blowout phase?

\item
Do all jets involve high-altitude magnetic reconnection occurring
at chromospheric and coronal null points, or can the reconnection
occur in other magnetic field configurations involving high
spatial magnetic field gradients but no null points per se?
\end{itemize}

Regarding the last question, it is indeed known that coronal
energetic events do not always involve three-dimensional null
points. For example, X-ray bright points \citep{1996Mandrini} and
compact flares \citep{1997Schmieder} can sometimes occur in
quasi-separatrix layers \citep[as defined
  in][]{1996Demoulin}. Also small flares \citep{1998Aulanier,
  2004Pariat}, chromospheric surges \citep{2002Mandrini},  and a
recently studied recurring jet \citep{2013Guo} have been related
to bald patches \citep[as defined in][]{1996Bungey}.

With those questions in mind, in this paper we study the
evolution and the magnetic field configuration of a peculiar
coronal jet that was observed by the Atmospheric Imaging Assembly
(AIA; \citealt{2012Lemen}) onboard the \textit{Solar Dynamics
  Observatory} (\textit{SDO}). This event was peculiar in many
respects. Firstly, \textit{SDO}/AIA extreme-ultraviolet (EUV)
observations revealed complex interleaved and broad emitting
features, both along the jet itself and near its numerous
footpoints. Secondly, the line-of-sight magnetic field observed
by the Helioseismic and Magnetic Imager (HMI;
\citealt{2012Scherrer,2012Schou}) onboard \textit{SDO} showed
that the jet was launched from a weak-field area   surrounded by 
strong network polarities resulting from  the decaying  phase of the  active region (AR11106).  The
jet itself cannot be unambiguously assigned to the quiescent or
blowout types, and there is no clear hint of ejection of a flux
rope as a mini-CME.  We present the observations and our data
analysis in Section 2. A nonlinear force-free field extrapolation
and topology analysis of the event is given in Section
3. Finally, in Section 4, we summarize and discuss the results in
the context of the aforementioned questions.

\section{Observations and data analysis}

\subsection{AIA and HMI instruments} \label{sec:instrument}

\begin{figure*}
\centering

\includegraphics[width=16cm]{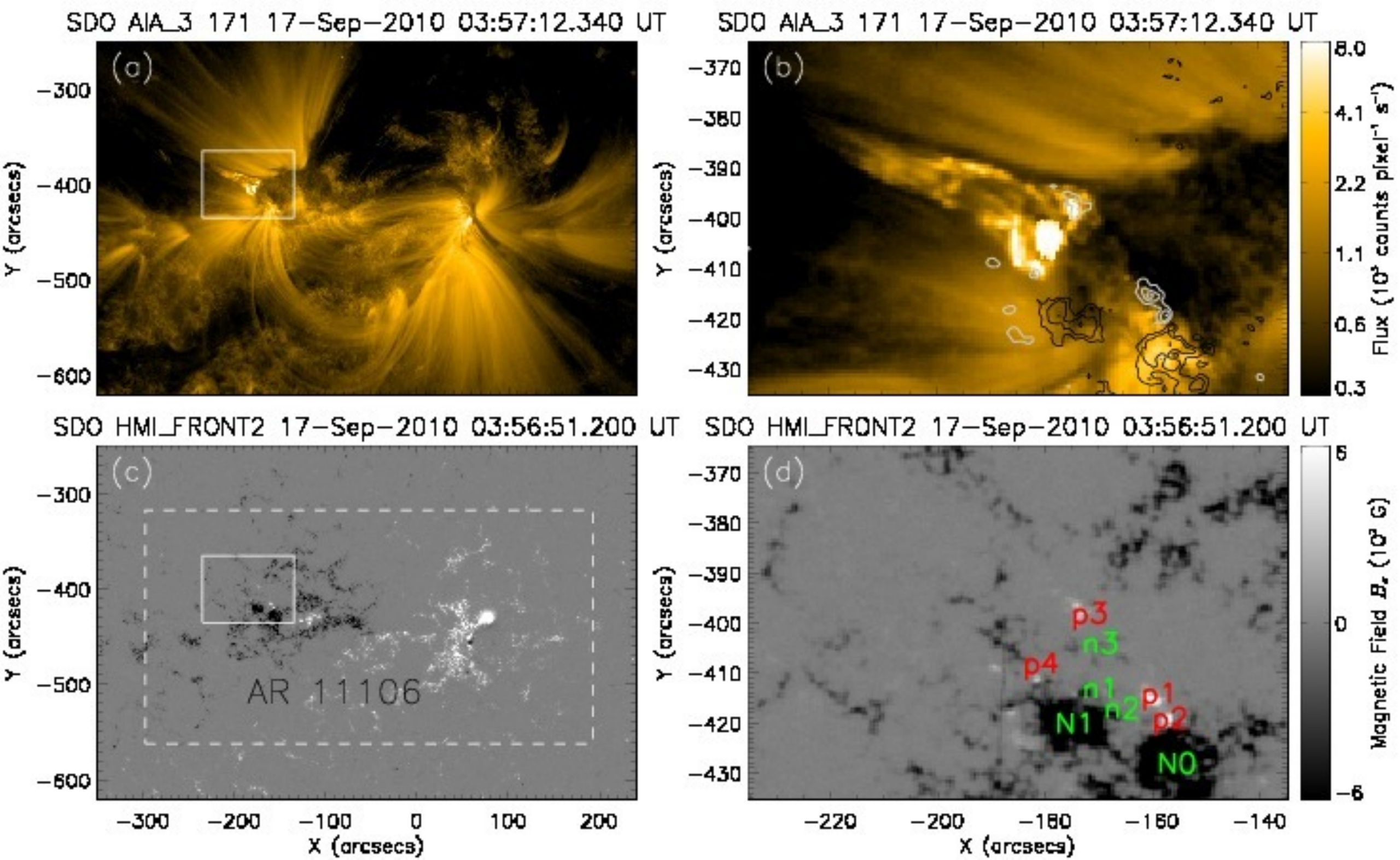}
\caption{\textit{SDO}/AIA 171~\AA \ image and \textit{SDO}/HMI
  line-of-sight magnetic field. \textbf{(a) } A jet in 171~\AA
  \ observed by \textit{SDO}/AIA at 03:57 UT. The solid box
  indicates the field of view of panels (b) and (d). \textbf{(b)}
  The same 171~\AA \ image as that in panel (a) but in a smaller
  field of view. Gray/black lines represent positive/negative
  polarities of the line-of-sight magnetic field as shown in
  panels (c) and (d). The contour levels of $B_\mathrm{LOS}$ are
  -1200, -900, -600, 50, 300, and  600 G. \textbf{(c)} Line-of-sight
  magnetic field of the active region observed by
  \textit{SDO}/HMI at 03:56 UT. The solid box is the same as 
  in panel (a). The dashed box indicates the field of view used
  for removing the 180$^\circ$ ambiguity and the projection
  effect. \textbf{(d)} The same line-of-sight magnetic field as
   in panel ©, but in a smaller field of view. Mixed
  parasitic polarities are denoted by p1, p2, p3, and p4 for
  positive polarities and n1, n2, and n3 for negative
  polarities. N0 and N1 denote the main negative polarities in
  the following part of the active region.} \label{fig:aiahmi}
\end{figure*}

\textit{SDO}/AIA provides filtergrams in seven EUV spectral lines
and three UV-visible continuums with high cadence (12 s) and high
spatial resolution ($1.5''$). Each of its four CCD arrays has
$4096 \times 4096$ pixels and the pixel sampling is $0.6''$. The
field of view is $41'$ x $41'$, hence fully including a disk of
radius $1.3\, R_\odot$.  Due to its large spectral coverage,
\textit{SDO}/AIA observes the solar atmosphere over a large
temperature range from the cold photosphere at about 5000K (white
light and 1700~\AA ) to the hot active corona at around 10MK
(94~\AA\ , 131~\AA\ and $193$~\AA ).

\textit{SDO}/HMI observes the polarimetric line profiles at Fe I
6173~\AA \ with a filter and two CCDs on the full disk of the
Sun. Each of the two CCDs has $4096 \times 4096$ pixels. The
spatial resolution is $1''$ with a pixel size of $0.5''$. As a
filtergraph, \textit{SDO}/HMI scans the spectral line profile at
six positions. The full width at half maximum (FWHM) of the
filter is 76~m\AA \ and the spacing of the filter position is
69~m\AA . The two CCDs record two sets of data for computing the
line-of-sight and vector magnetic field, respectively. For the
first case, only the $I \pm V$ ($I$ and $V$ stands for the Stokes
parameters) filtergrams at the six wavelengths are recorded, and
the cadence is about 45 s. For the second case, four additional
filtergrams ($I \pm Q$ and $I \pm U$, where $Q$ and $U$ stand for
the other two Stokes parameters) at each of the six wavelengths
are recorded, and the cadence is about 135 s. The Stokes
parameters $I, Q, U$ and $V$ are computed from the raw data after
necessary calibrations. However,  to increase the signal-to-noise ratio, the Stokes parameters are averaged over 12
minutes. The vector magnetic fields are derived using the code  {\it
  very fast inversion of the Stokes vector}  (VFISV;
\citealt{2011Borrero}).

\subsection{EUV jet and line-of-sight magnetic field}\label{sec:euvjet}

\begin{figure*}
\centering
\includegraphics[width=1.\textwidth]{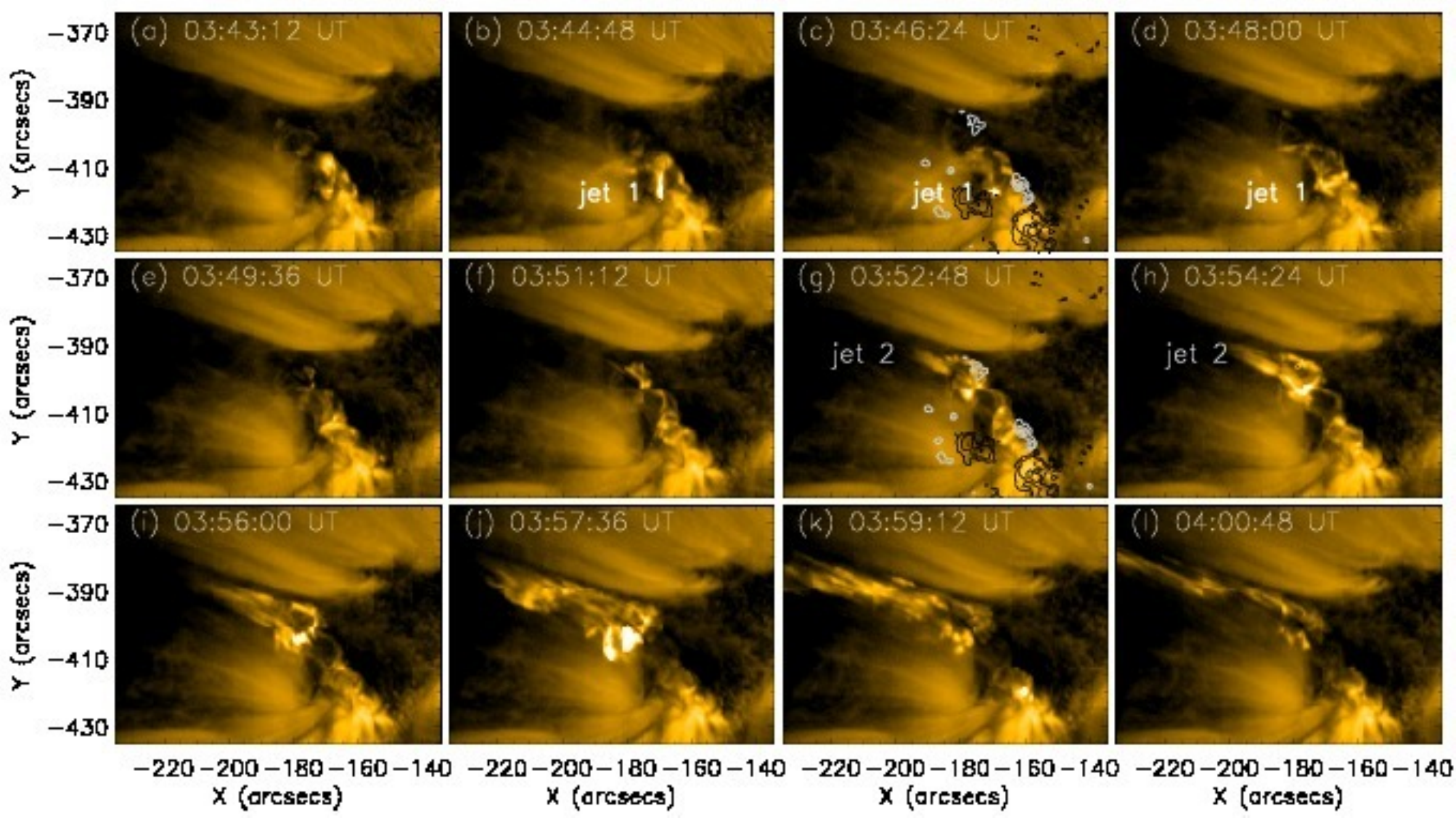}
\caption{\textit{SDO}/AIA 171~\AA \ images showing the evolution
  of two EUV jets. The color scale is the same as in
  \fig{fig:aiahmi}. The labels jet 1 and jet 2 denote the
  two jets close to each other both in time and space. Grey/black
  lines in panels (c) and (g) represent positive/negative
  polarities of the line-of-sight magnetic field at 03:56:51
  UT. The contour levels of $B_\mathrm{LOS}$ are -1200, -900,
  -600, 50, 300, and 600 G. } \label{fig:aiaevo}
\end{figure*}

Many bright points and jets (around 30) were observed in the
eastern and western edges of active region (AR) 11106 on
September 17, 2010. We focus our study on the main jets occurring
in the following (eastern) polarity of AR 11106, where
\citet{2013Guo} studied three major recurring jets, whose SDO/AIA
171~\AA \ flux peaked at 03:17, 03:57  and 04:22 UT,
respectively. From Figures~2 and 5 of \citet{2013Guo}, they found
that the 171~\AA \ flux curve around 03:57 UT shows some smaller
peaks before the main peak, while the other two events at 03:17
and 04:22 UT have only one peak.  The 03:57-UT jet is labeled
\jettwo\  in the following while the recurrent jets studied
by \citet{2013Guo} are collectively called \jetone. One of
those jets occurs close in space and time to \jettwo, but the
parasitic polarities involved in \jetone\ and \jettwo\ are
different, as we show below.   The detailed
evolution and magnetic configuration of the peculiar event at
03:57 UT we call \jettwo\ have not been studied yet and
constitute the object of the present
paper. Figure~\ref{fig:aiahmi}a shows large-scale coronal loops
connecting the two main polarities in AR 11106 observed with
\textit{SDO}/AIA 171~\AA . A coronal jet is present at the east
(left) edge of AR 11106. The footpoints of the jet are located at
some parasitic positive polarities, as shown in
Figure~\ref{fig:aiahmi}b and d.

AR 11106 survived for several solar rotations. It mainly
consisted of  a dispersed magnetic field on each side of a large
filament channel, which can be seen as the dark lane in
Figure~\ref{fig:aiahmi}a below the large-scale coronal
loops. This active region was observed in August, September,
October, and November in 2010. In September, there was still a
leading positive polarity spot and the following polarity was
mainly formed by intense network polarities of kilo Gauss
strength (Figure~\ref{fig:aiahmi}c). Inside the following
negative polarity, emerged  continuously  bipoles. The
main emergence occurred on 2010 November 9, when a new active
region was formed inside the remnant region. On 2010 September
17, there were two locations of positive polarities in the
following negative polarity region in the intra-network
(Figure~\ref{fig:aiahmi}c). We focus on the northeast one 
shown in the solid box of Figure~\ref{fig:aiahmi}c. The polarities
correspond to the emergence of bipoles occurring one or two days
before. On September 17, the active region was close to the
central meridian.

In Figure~\ref{fig:aiahmi}d, we present a small field of view
concentrating on the following part of the active region with two
main polarities labeled N0 and N1 ($B_\mathrm{LOS} > 1000$
G). Weaker mixed polarities ($B_\mathrm{LOS}$ of  about 100
to 600~G) are inside a supergranule  identified in the dispersed 
magnetic field of the active region.  It is surrounded by negative polarities (N0 and
N1 to  the nouth of the supergranule). There are many parasitic polarity regions corresponding to
different systems of emerging flux, for example, p1--n1, p2--n2,
and p3--n3.

\begin{figure}
\includegraphics[width=0.48\textwidth]{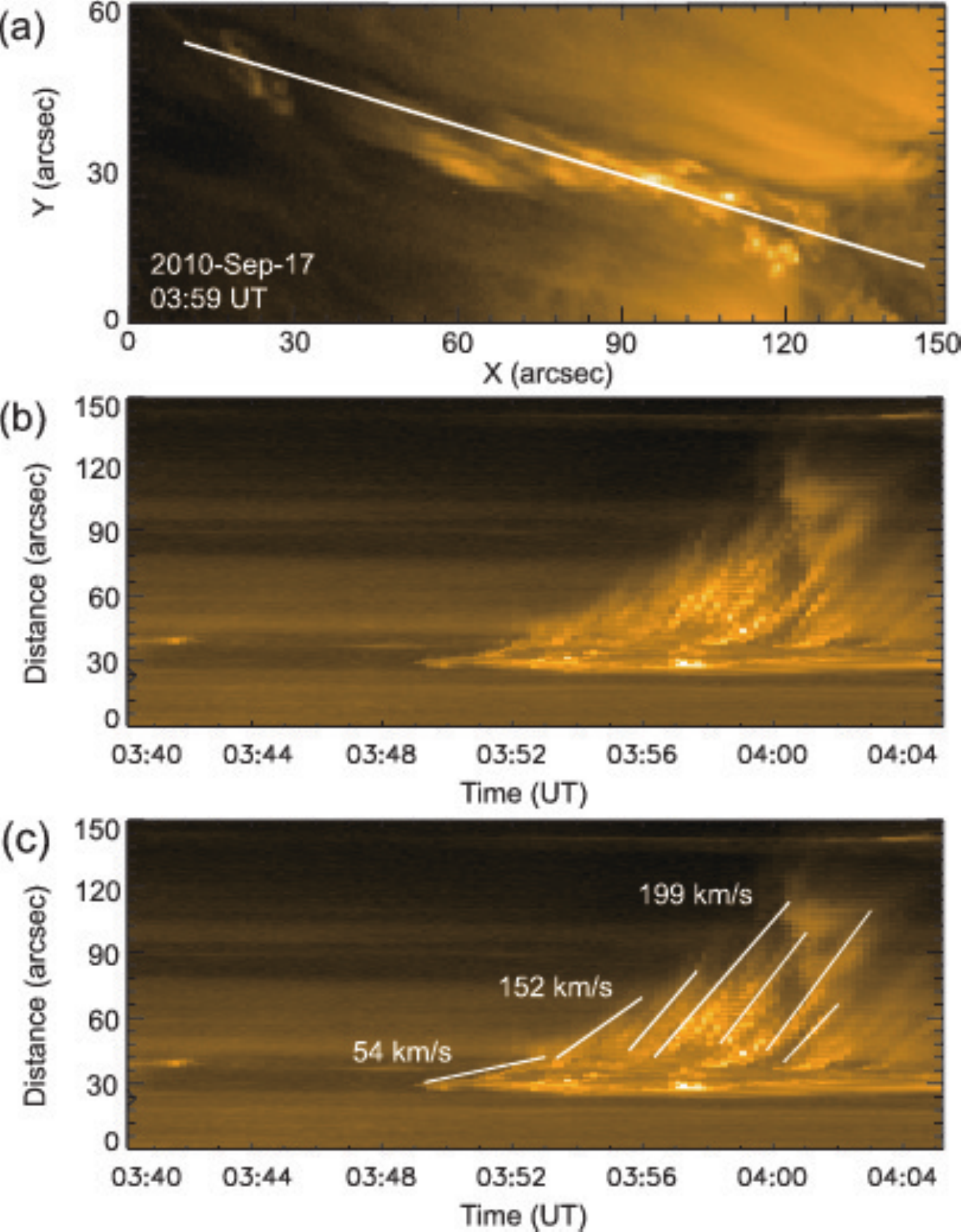}
\caption{\textbf{(a)}  EUV jet observed with the 171 \AA
  \ filter of \textit{SDO}/AIA  on 17 September
  2010. The solid line indicates the slit position along which
  the jet was ejected. \textbf{(b)} Temporal evolution of the
  171~\AA \ image in the slice along the jet.  \textbf{(c)} Some
  linear fitted lines overlaid on the time-slice 171 \AA
  \ image. As  time passes, different components or
  kernels of the jet were launched from the jet base with 
  increasing speed, reaching  200 km~s$^{-1}$.
} \label{fig:slit}
\end{figure}

\fig{fig:aiaevo} shows the evolution of the EUV jets in 171~\AA
\ from 03:43 to 04:00~UT. Only bright points can  be seen at
03:43 UT,  as shown in \fig{fig:aiaevo}a. At 03:44 UT, a small
curved jet (\fig{fig:aiaevo}b through d) appeared above the
parasitic polarities p1--n1 and p2--n2. We labeled this \jetone. It extended toward the southeast and
lasted until 03:48 UT. At 03:51 UT (\fig{fig:aiaevo}f), a small
jet emanating from p3 began to the north of \jetone. The jet
(\jettwo) extended toward the east with some brightening kernels
at its footpoint in the west end (\fig{fig:aiaevo}g). Later on
(Figures~\ref{fig:aiaevo}h, \ref{fig:aiaevo}i, and
\ref{fig:aiaevo}j), we find many threads along the jet and some
brighter kernels at its basis. At 03:59~UT (\fig{fig:aiaevo}k),
the jet was composed of about ten threads with an angle of about
10 to 20 degrees aligned in  the direction of the ejection. The
global shape of the jet looks like the  Eiffel Tower and  it
consists of different branches with seemingly torsional
threads. The number of bright threads decreased at 04:00 UT
(\fig{fig:aiaevo}l) while the remaining long thread  still
increased in size and speed. Jet 2 disappeared at about 04:08
UT.

Jet 2 was also clearly observed in other filters of
\textit{SDO}/AIA, for example, 304~\AA , 193~\AA , 221~\AA , and
1600~\AA. The jet had many  thermal components that not
completely co-spatial. Between 03:34 to 03: 41 UT, the jet was
visible as a dark structure in absorption in the 304~\AA \ filter,
which indicates a surge-like behavior. Some brightenings were
detected at the jet base in 1600 \AA . The jet was also observed
in soft X-rays with Hinode/XRT.

\subsection{Time-Slice analysis of the EUV jet}\label{sec:slice}

\begin{figure}
\centering
\includegraphics[width=0.48\textwidth]{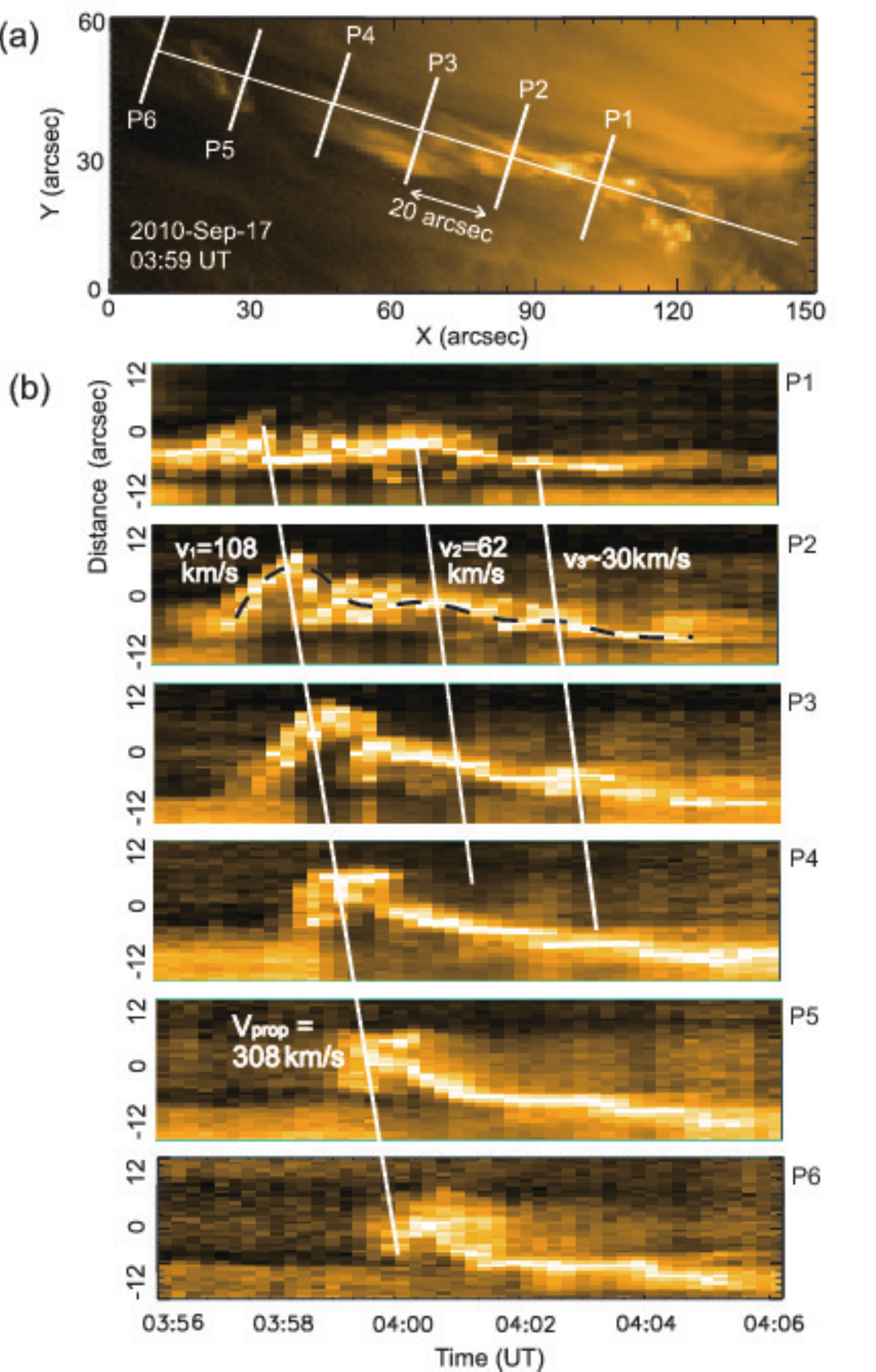}
\caption{\textbf{(a)} EUV jet observed with the 171 \AA
  \ filter of \textit{SDO}/AIA  on 17 September
  2010. The shorter solid lines indicate the slit positions at six
  different heights used for (b), which are perpendicular to the
  longer solid line. The length of the slits is
  $24''$. \textbf{(b)} Time-distance diagram of the 171 \AA \ EUV
  jet along the slits at the six different heights. The dashed black line
  shows the oscillation pattern at one height, and the white lines
  indicate the propagation velocity of the oscillation patterns
  at  different heights.  } \label{fig:slit2}
\end{figure}

To quantify the propagation of the jet, we made a time-slice
analysis. First, we selected a straight line (called a slit in the
following) roughly along the jet on the 171-\AA~SDO/AIA image, as
shown in \fig{fig:slit}a. Next, we drew the brightness
distribution along the slit for successive times with the aim  to
discern the propagation of features in the jet (\fig{fig:slit}b,
c). In the figure we see that some bright features started to
rise at around 03:49~UT with a speed of 54 km s$^{-1}$. Presently, a striped pattern of dark and light lines or strands
becomes apparent; the pattern extends to higher values of the
ordinates as time passes. The pattern suggests the propagation
of a wave motion; it would be compatible with the signal left by
untwisting field lines carrying the hot material, or, more
generally, with an Alfv\'en wave propagating along twisted field
lines.  However, some other mechanisms would also account for it,
such as intermittent ejections of plasmoids along straight field
lines or successive plasma ejections along sheared field
lines. Moreover, because the 171~\AA \ emission is optically thin, it
is impossible to determine the chirality of any twist of the
field lines involved in this emission pattern (see also the
discussion in Section~\ref{sec:topo}).  As time passes,
different components or kernels of the jet are launched from the
base with increasing  speeds, reaching  in projection  $200$ km s$^{-1}$.  They
reach a projected length of 100 Mm. This indicates that the EUV
jet was becoming more dynamical during this phase of its life,
perhaps as a result of the relaxation of a twisted structure.
This speed is comparable with the results of other authors
\citep{2007Cirtain,2008Patsourakos}.  As said in
Sec.~\ref{sec:euvjet}, the jet was also observed with Hinode/XRT
(with temperature of around a few million degrees). The spatial
resolution is lower and the determination of the jet speed has a
large error. Nevertheless, the increase in velocity was also
observed.

To study the evolution of the jet plasma that is perpendicular to
the propagation direction, we made a fishbone-slit cut as shown
in \fig{fig:slit2}a.  Six slits were selected perpendicular to the
jet propagation direction. The distance between each pair of
adjacent slits is $20''$. The 171~\AA \ intensity distribution
along each of the six slits is arranged as a function of time in
\fig{fig:slit2}b, with panels labeled P1 through P6. Any apparent
vertical motion in these panels thus corresponds to sideways
elongation of the bright features in \fig{fig:slit2}a away from
the main axis of the jet.  In each of the six slits, we find that
the bright features moved basically along the whole length of the
P1 - P6 transverse slits. We also discern a time shift between
the patterns seen in successive panels, from P1 through P6: we
have drawn slanted white straight lines to indicate this
shift. Additionally, there is an apparent swing or oscillation in
the motion of the main bright feature in  panels P1 - P3: we
have drawn a solid dashed black curve in P2 to highlight this apparent
phase motion. The dashed dark  curve seems to indicate a damped
oscillation taking place sideways of the main jet axis. We found
three periods of oscillation and the numbers above the curve
denote the amplitude of the corresponding velocity, namely $108$
km s$^{-1}$, $62$ km s$^{-1}$,  and $30$ km s$^{-1}$, respectively, 
with an error estimated to be around 10 km/s.
This may indicate that the oscillation is triggered in the
initial phase, slowly decaying thereafter.

\begin{figure}
\includegraphics[width=0.48\textwidth]{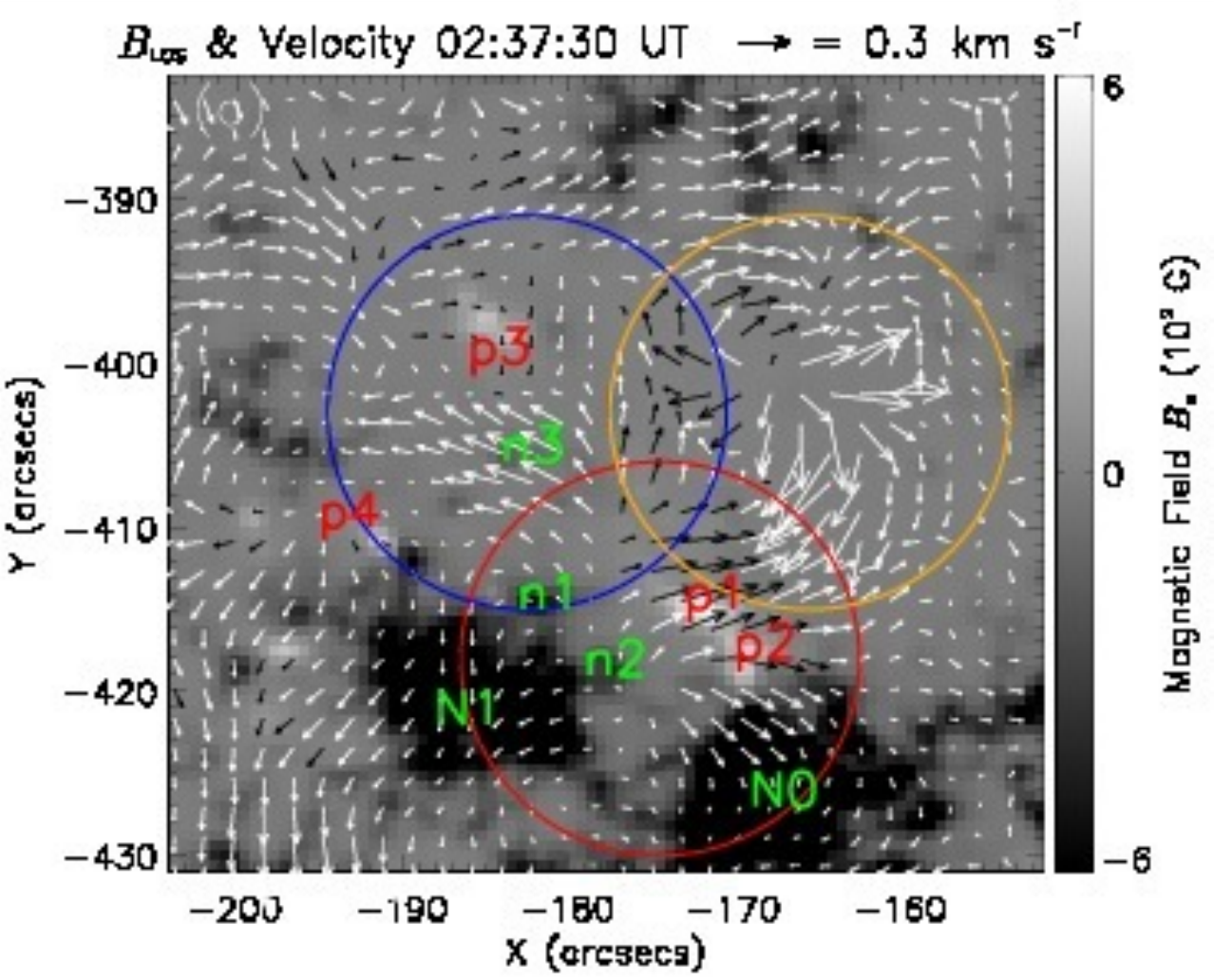}
\caption{Photospheric transverse velocity derived through
  LCT analysis of the \textit{SDO}/HMI magnetograms. The
  background is the average line-of-sight magnetic field with
  positive (white) and negative (black) polarities. Arrows
  represent the transverse velocity, where black/white is located
  on the positive/negative polarity. The time interval for
  averaging the line-of-sight magnetic field is 02:25--02:50~UT.
} \label{fig:lct}
\end{figure}

\subsection{Evolution of the line-of-sight magnetic field}

To study the transverse flows of photospheric magnetic features,
we employed the local correlation tracking technique (LCT;
\citealt{1988November}). A Gaussian tracking window with a full
width at half maximum (FWHM) of $5''$ was adopted. A time series
of 137 \textit{SDO}/HMI magnetograms starting at 02:00 UT on 2010
September 17 and with a cadence of 45 seconds was used for the
analysis. All the magnetograms were aligned with each other prior
to the LCT analysis. We refer to \citet{2013Guo} for detailed
descriptions on the alignment. Then, the magnetograms were grouped
and averaged in $25$-minute series to decrease the noise. The
transverse velocity maps were derived by the LCT technique with
the averaged magnetograms. They were finally trimmed to include
only the area  present in all frames.

\fig{fig:lct} contains one of the LCT velocity maps obtained in
that process. The figure shows strong diverging and converging
velocity distributions. We selecedt three regions in the map
(indicated with circles in red (bottom), orange (top right), and blue (top left) ) to investigate
the flow patterns in detail.  In the region enclosed by the red
circle, the weak polarities p1--n1 and p2--n2 continuously
emerged in the intra-network and moved at a speed of $\sim 0.2$
km~s$^{-1}$ toward the preexisting main polarities N0 and N1 at
the edge of the supergranule. The opposite polarities within each
of the pairs p1--n1 and p2--n2 move-away from each other, which
suggests they were emerging bipoles. The magnetic polarity p2
moves toward N0. This polarity pattern has been interpreted as
an instance of emerging flux causing the jet series called
\jetone\ \citep{2013Guo}.

For the region enclosed by the blue circle (top left) toward the center of
\fig{fig:lct}, we see a cluster of negative polarities, which we
call n3, moving toward the northeast (upper left), in the
direction of  polarities p3 and p4.  This is the location from
which jet 2 is launched.

In the third region, indicated in \fig{fig:lct} by an orange
circle (top right), a diverging flow pattern with a clockwise
rotation is apparent. The velocity vectors are discontinuous in
the common area of this circle with either the blue or the red
ones. The velocity arrows near the
orange-red intersection are directly opposite, and  we see in the
orange-blue one a rotational shear.  We show in Sec.~\ref{sec:topology} that
there are bald patches in these intersections and, generally, in
the interior of the orange circle; there is also an extended
bald patch to the north-east of the blue circle.

\begin{figure*}
\centering
\includegraphics[width=0.75\textwidth]{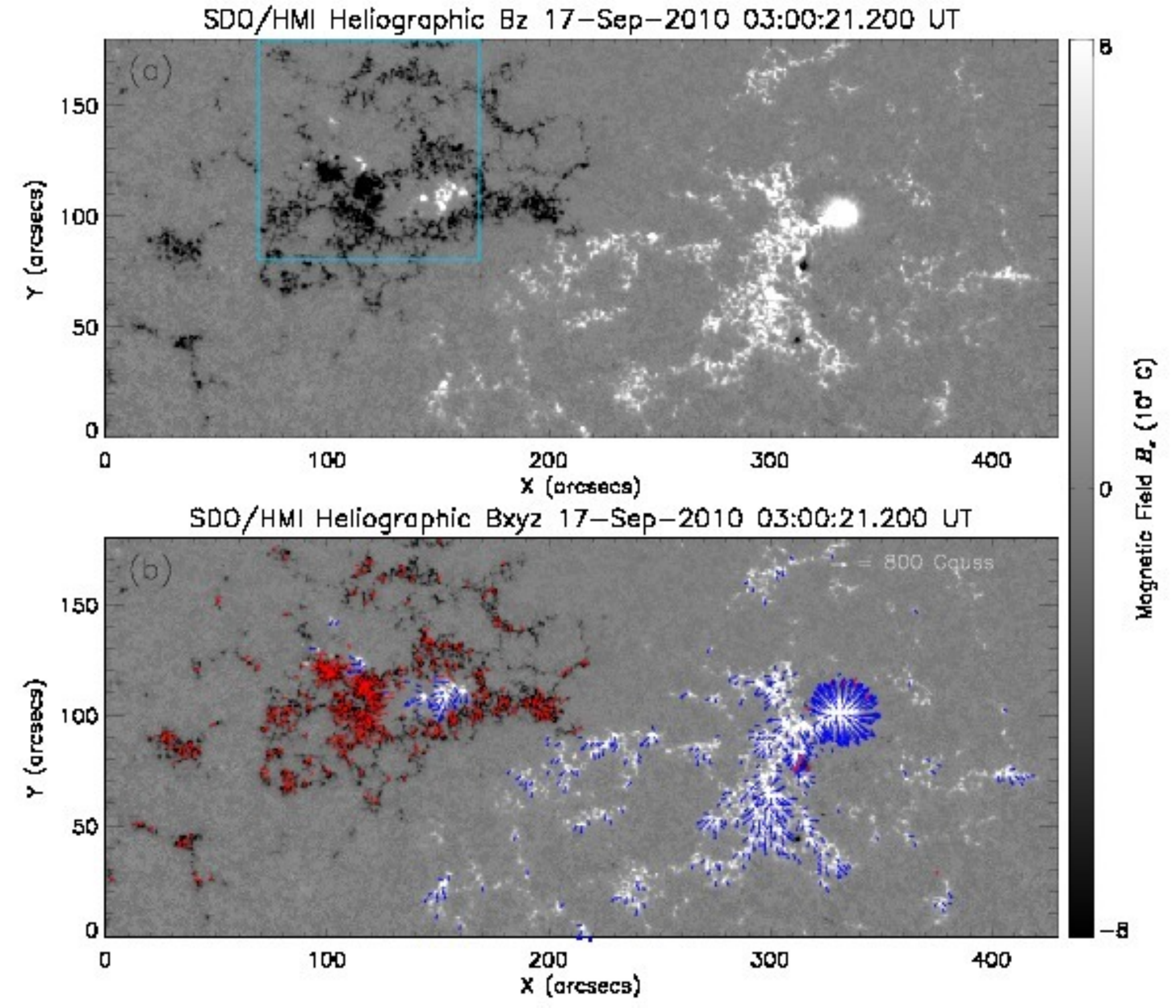}
\caption{\textbf{(a)} Grayscale map of the vertical component of
  the vector magnetic field in the domain used for the NLFFF
  extrapolation (the field strength is indicated in the vertical
  bar on the right). The field  strength shown in the figure  is corrected from the  projection
  effects  but has not yet been subjected to
  preprocessing.  The solid box marks the region used in later
  sections to compute the bald patches and to plot magnetic field
  lines. \textbf{(b)} Vertical and horizontal components of the
  vector magnetic field. The direction and length of the arrows
  indicate the direction and strength of the horizontal magnetic
  field, respectively. For the conversion from arrow length to
  field strength, a sample arrow is shown in the upper right
  corner.} \label{fig:bxyz}
\end{figure*}

\section{NLFFF extrapolation and topology analysis} \label{sec:topo}

\subsection{Vector magnetic field and NLFFF extrapolation}

For the extrapolation, we used the photospheric vector magnetic
field data obtained through SDO/HMI as described in
Section~\ref{sec:instrument}. For the following analysis  we
used the data corresponding to 03:00 UT.  The transverse
components of the vector magnetic field suffer from a $180^\circ$
ambiguity, which was removed with  the improved version of the
minimum-energy method
\citep{1994Metcalf,2006Metcalf,2009Leka}. Then, we corrected for
the projection effects with  the method proposed by
\citet{1990Gary}. The correction was carried out in two steps.
First, the line-of-sight and transverse components of the vector
magnetic field were transformed to the heliographic
components. Then, these fields were projected from the image plane
on to the plane tangential to the solar surface at the center of the
selected field of view. In this case, the center is located at $x
= -60.0''$ and $y = -439.5''$ at 03:00 UT. Correction of the
projection effects is necessary since the region of interest is
located at a relatively high latitude in the southern
hemisphere. After the correction, any fake polarities associated
with the projection disappear. The corrected vector magnetic
field is shown in Figure~\ref{fig:bxyz}.

To derive the three-dimensional magnetic field configuration, we
 extrapolated  the photospheric boundary data
assuming the field to be force-free, that is  by solving the
equations $\nabla \times \mathbf{B} = \alpha \mathbf{B}$ and
$\nabla \cdot \mathbf{B} = 0$. The pseudo-scalar $\alpha$ must be
constant along individual field lines (as follows from 
combining the  two equations), but otherwise needs not be
constant in space.  The de-projected vector magnetic field was
used as the bottom boundary condition (bc); the boundary conditions on the other
boundaries of the cubic box were set by using the results of a potential
field extrapolation of the vertical component of the photospheric boundary data.  The
resulting problem is nonlinear, and we select the optimization
method \citep{2000Wheatland,2004Wiegelmann,2010Wiegelmann} to
obtain the solution.

A requirement for successful NLFFF modeling is that the net
magnetic flux on the photospheric boundary must be near zero,
that is  the flux of the positive and negative magnetic patches in
the boundary data approximately balance. In practice, this
condition requires that the photospheric vector data used for the
NLFF modeling include all  strong positive and
negative magnetic flux concentrations  of the AR, and that the latter are
sufficiently far from the boundaries of the selected subfield
\citep{2009DeRosa}. Only then do the field lines issuing from the
strongest flux and electric current concentrations reconnect 
to the lower boundary of the NLFF model field and one can reasonably
assume to have properly captured the magnetic connectivity within
the active region. For our de-projected observational boundary
data (shown in Figure~\ref{fig:bxyz}), the flux-balance
parameter, which is defined as the absolute value of the ratio of
the net signed vertical magnetic flux over the total unsigned
vertical magnetic flux, is $0.034$,  that is,  it is nearly flux
balanced.

In addition, the force-free assumption for the magnetic
field in a given volume leads to specific conditions for the field on the
boundary of the volume related to the Lorentz force itself and to its torque
\citep{1989Aly}; in our case, these conditions are unlikely to be satisfied,
not even approximately, by the observational boundary data, -specially because 
the plasma $\beta$ is not low at the photosphere \citep{1995Metcalf,
  2012Tiwari}. To solve this problem, \citet{2006Wiegelmann} proposed a
preprocessing of the boundary data. The method has four input parameters,
$\mu_1, \mu_2, \mu_3$, and $\mu_4$, which regulate the quality of the boundary
condition resulting from the preprocessing concerning not only the zero-force
and -torque conditions, but also the closeness to the observed data and the
smoothing. 
The method  leads to a decrease of the dimensionless magnetic force
parameter from $0.420$ to $0.001$ before and after the preprocessing,
respectively, and of the dimensionless torque parameter from $0.400$ to
$0.002$. As noted by \citet{2006Wiegelmann}, vector maps with
dimensionless force and torque parameters on the order of $10^{-2}$ can
validly be used as input for the NLFFF magnetic modeling.
The flux balance parameter becomes $0.038$ due to the preprocessing, that is, it remains equally
low.

The preprocessed vector magnetic fields were then used as boundary input in
the improved optimization method of \citet{2010Wiegelmann} to obtain a
solution of the NLFFF problem.  The improved algorithm relaxes the magnetic
field not only in the volume but also in the lower boundary of the cubic
computational domain. We followed the prescriptions provided by Wiegelmann et
al. (2012) concerning the weighting and injection speed of the lower-boundary
data.  The extrapolation was performed in a cubic box  resolved by $860
\times 360 \times 680$ grid points. As an indication of the quality of the
magnetic field obtained, we mention that the average angle between the
magnetic field and the electric current density  is about
  $10.6^\circ$.

\subsection{Magnetic topology of \jettwo}\label{sec:topology}

Coronal jets are widely accepted  to be  the consequence of magnetic
reconnection, which often  occurs in magnetic structures with
singular topologies, such as null-points with their associated
fan-spine configurations \citep{2011AdSpR..47.1508P}.  On the
other hand, magnetic reconnection is also possible in
Quasi-Separatrix-Layers (QSLs), which, in fact, may account for
the launching of the jet complex (\jetone) studied by
\citet{2013Guo}. QSLs are likely to appear associated with bald
patches (as discussed below). To understand the structure of the
jet analyzed in the present paper (Jet 2), we therefore aimed at 
locating the possible null points and bald patches in the area of
interest where the jet has its roots.

\begin{figure}
\includegraphics[width=0.4\textwidth]{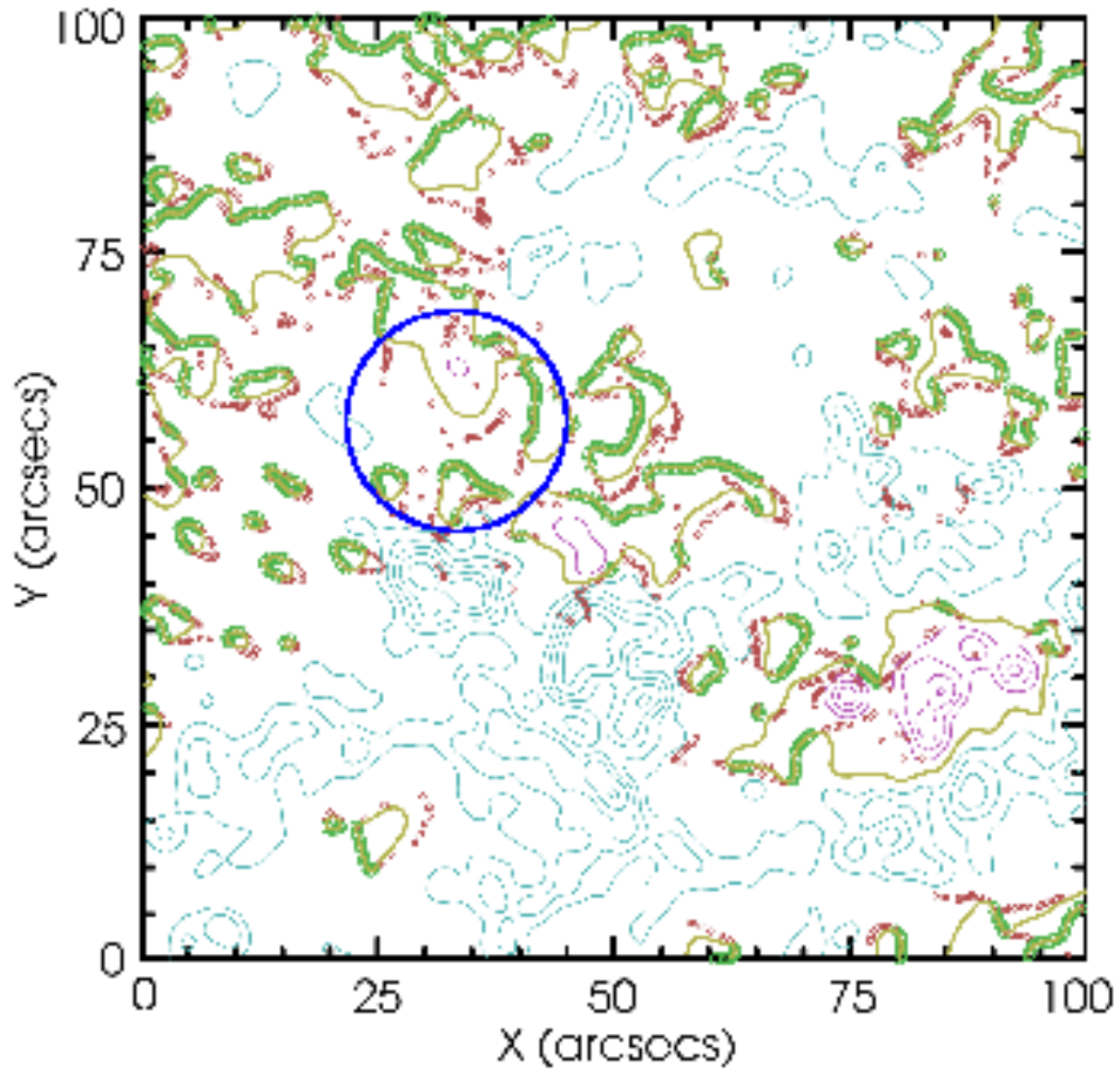}
\caption{Bald patches (green circles) computed from the NLFFF
  field model. Dark red circles represent the footpoints of the
  bald patch field lines. Magenta/cyan solid lines denote the
  contours of the positive/negative polarity. The olive-green
  solid line indicates the polarity inversion line, where all the
  bald patches are located. The blue circle in this figure is the
  same as shown in \fig{fig:lct} and marks the region from where
  \jettwo\ is launched.} \label{fig:bald}
\end{figure}

Magnetic dips are magnetic field lines that are concave in their upper part. The
magnetic field at the bottom of magnetic dips fulfills the
conditions $\mathbf{B} \cdot \nabla B_z > 0$ and $B_z = 0$, with
$z$ the upward-pointing vertical coordinate.  Bald patches are
magnetic dips whose bottom touches the photosphere. With this 
definition, the bald patches were computed and  are shown by the small
green circles in Figure~\ref{fig:bald}. They all lie along the
polarity inversion line, shown as a thin olive-green line.  The
small dark-red circles in the figure mark the foot points of the
field lines that are tangential to the bald patches.  
To facilitate identification, we have repeated in the figure the blue circle
of \fig{fig:lct}. To the west (i.e., right) of the blue circle,
we see a collection of bald-patch lines associated with the p1
and p2 positive polarities located above the strong  negative
polarity labeled N0.  These bald patches are situated in the
interior of the orange circle of \fig{fig:lct}.  The closeness in
space of these bald-patch lines and their different orientations
reflect how  complicated  the field line structure is in that
region. We  traced the magnetic connectivity of these bald
patches using the NLFFF extrapolation data. Most of the field
lines going through them just end up connecting to the main
positive polarity of the active region via simple large-scale
loops. However, as one moves in the region toward the north,
field lines connecting to the roots of the jet begin to
appear. The bald-patch region near the right boundary of (but
still inside) the blue circle is particularly interesting: this
is the region of high-velocity shear in \fig{fig:lct}; the field
lines traced from that region connect in part to the jet area and
in part to the positive AR polarity, with strong field-orientation changes. It is clear that important  field-line shear
is taking place there.

Back in \fig{fig:bald}, we see another extended  bald-patch line
to the north-east (i.e., top-left) of the blue circle. These bald
patches are associated with the weak positive polarity p3 (which
appears as a small magenta ring within the circle). We show
presently that this region has a complicated and interesting
topology, with very different large-scale connectivity patterns
for neighboring patches, which can also explain part of the
reconnection that causes \jettwo.

\begin{figure*}
\centering
\centerline{\includegraphics[width=0.7\textwidth]{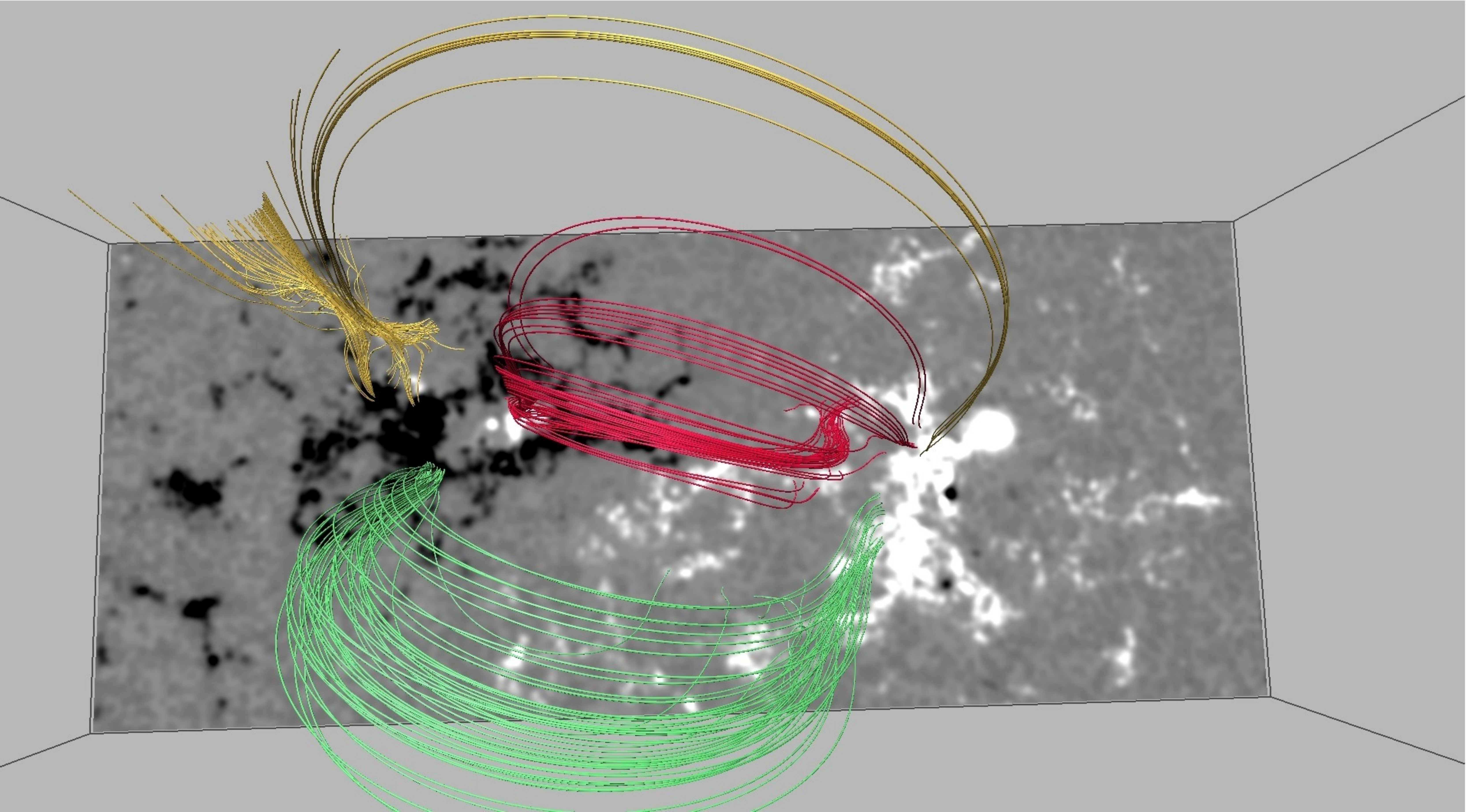}}
\vskip 1mm
\centerline{\hfill
\includegraphics[height=0.35\textwidth,width=0.08\textwidth]{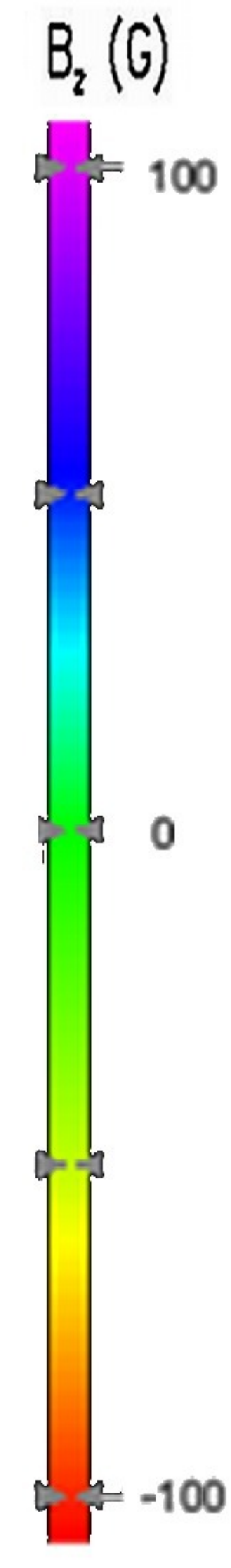}\hskip 1mm
 \includegraphics[width=0.58\textwidth]{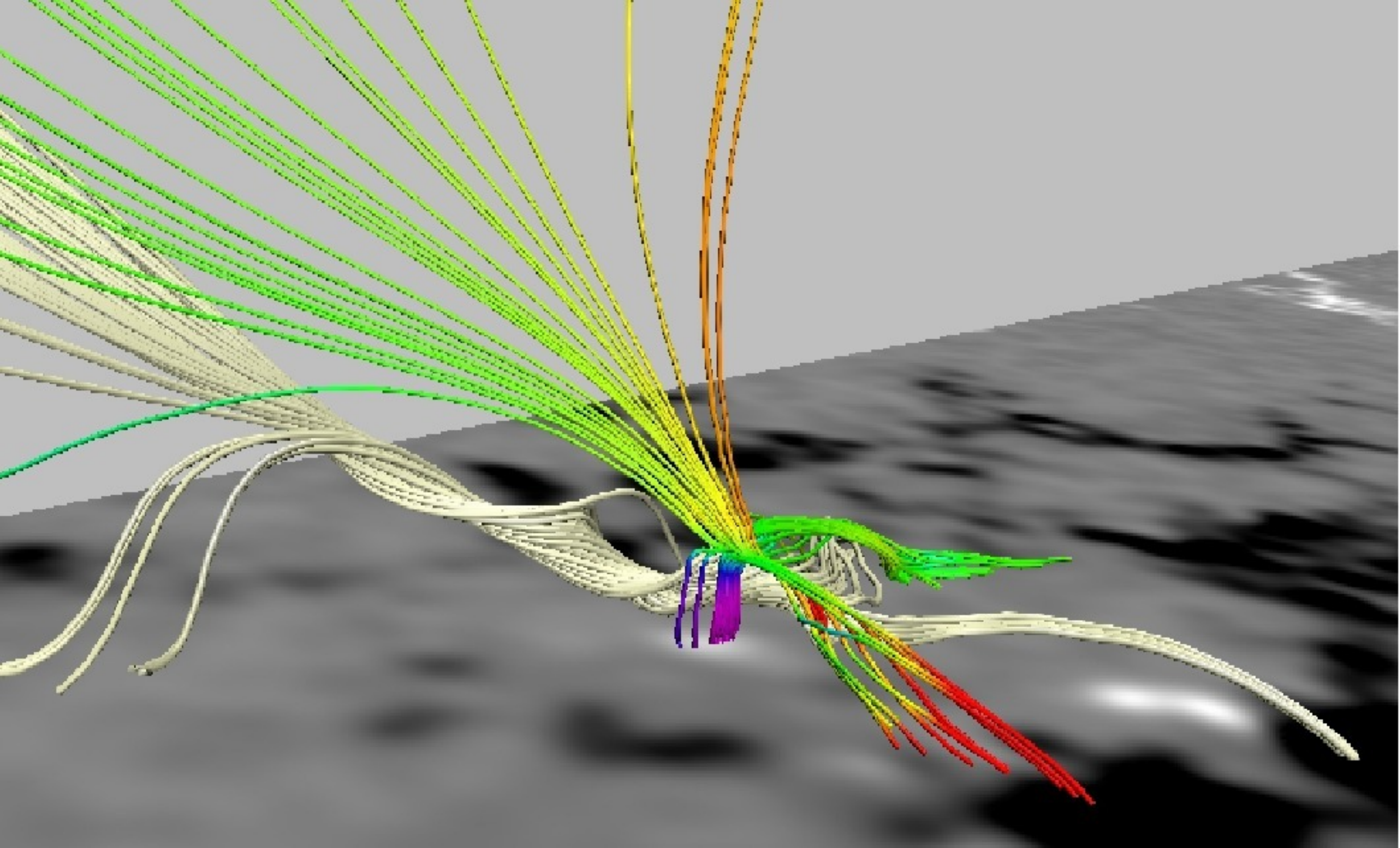}\hskip 1mm\hfill
 \includegraphics[width=0.3\textwidth]{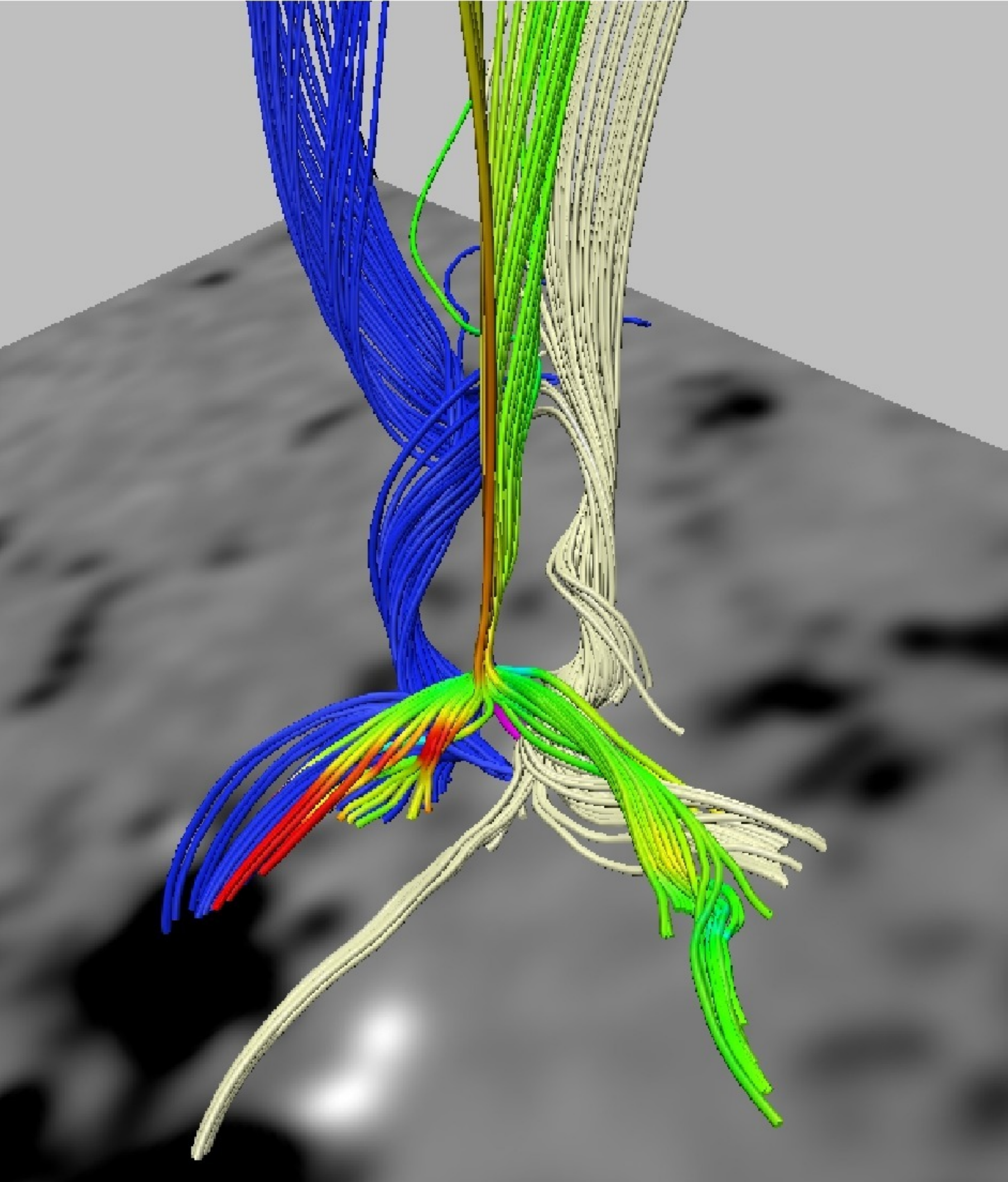}\hfill}

\caption{Field line connectivities obtained through the NLFFF
  extrapolation. Top panel: the green and red field line sets
  illustrate the general connectivity between the main polarities
  of AR11106 and resemble structures visible in the SDO/AIA image
  of \fig{fig:aiahmi}. The yellow field lines are traced from the
  region at the base of the jet and clearly reproduce  the geometry
  of the latter. Lower panels: local and remote connectivities of
  the region around the minor positive polarity p3. One of the
  field line sets is colored according to the value of $B_z$, the 
  colorscale follows the color bar on the left. The ivory and
   royal-blue field line sets show twisted flux ropes on either side of
  the central spine. } \label{fig:vis1}
\end{figure*}


\fig{fig:vis1} (top panel) illustrates the large-scale connectivity pattern
of various patches on the negative side of the active region, including the
supergranule on the north-east (top-left) that contains the positive
polarities p1-p2 and p3 (compare this figure with \fig{fig:aiahmi}; the
supergranule is contained in the small rectangle of panels a and c in that
figure). The coronal loops drawn in green and red in the top panel of
\fig{fig:vis1} roughly correspond to bright SDO/AIA loops seen to be rooted
toward the lower right corner of the rectangle in \fig{fig:aiahmi}a. On the
other hand, at the top-left corner of \fig{fig:vis1} (still in the top panel)
we see a complex field-line connectivity pattern above polarity p3 (yellow
field lines) that roughly coincides in location, shape, and orientation with
the complex bright structure at the base of jet 2 visible in
\fig{fig:aiahmi}a and b.  Closer inspection (\fig{fig:vis1}, bottom-left
panel) illustrates some of the topological features of the field above the p3
polarity. The panel shows two sets of field lines; in one of them the color
of the line varies with the sign and magnitude of the vertical component of
the magnetic field following the rainbow colors, as given in the color bar on
the left. Color saturation is reached at $|B_z|=100$ G.  This set of field
lines reveals a
singular structure similar to a highly asymmetric null point at a height of some $3.5$ Mm above p3.
 The
incoming field lines include a spine-like axis linking the structure to p3 on
the surface (purple field lines descending  to the white patch on the
surface) and, on the top side, field lines that come from the high levels of
the box (colored in pale green or blue). The outgoing field lines are issued
almost horizontally and link the structure to the negative polarities at the
periphery of the supergranule. At the center of this structure  is a
deep minimum of the magnetic field strength,  which can be identified
  with a null point within the accuracy of the NLFFF extrapolation. A linear
  analysis of the magnetic field structure around that point reveals three
  real eigenvalues of the field gradient matrix, one of which is much
  lower in absolute value than the other two: this confirms the highly
  asymmetric character of the null. There are more  structures of
interest in this domain: the vertical spine axis linking downward to p3 is
flanked on its NW and SE sides by two highly inclined (i.e., not far from
horizontal) flux ropes. One of them (the one to the NW) is shown in the panel
with field lines in ivory: this rope has a left-handed twist. The other
flux rope has field lines with a right-handed twist and is located to the SE
of the spine; it can be seen (deep-blue field lines) in the bottom-right
panel of \fig{fig:vis1}, which shows that site from a different
perspective. The two flux ropes extend toward the north-east of the FOV
roughly in the direction of \jettwo.

The topological relation of the structures just described in the neighborhood
of  polarity p3 with the bald patches to the north-east of it is
illustrated in \fig{fig:bald_patch_and_visualization}. The horizontal plane
in both panels contains a superposition of the NLFFF magnetogram (as also
used in \fig{fig:vis1}) and the diagram of \fig{fig:bald}, including the blue
circle: the two panels contain the same field line sets, viewed from
different perspectives. Three field line bunches have been drawn (in yellow,
gray and light magenta) that become tangential to the photosphere approximately at the
location of different bald patches to the north-east of p3. For better
identification with the previous figure, the set of field lines around the
singular structure visible in \fig{fig:vis1} is also repeated here, with the
same color-code along them.  

Combining the information of Figures \ref{fig:vis1} and
\ref{fig:bald_patch_and_visualization}, we conclude that \jettwo\ is launched
from a region at the edge of AR 11106, where strong gradients of
connectivity (QSLs) appear: patches connected with the strong positive polarity of
the active region sit next to domains which connect with the open field
outside of it. Moreover, the intrusion of positive polarities into a supergranule
with the predominant polarity of that side of the active region causes a
collection of bald patches  and a very asymmetric
    null point to appear. This is clearly a region well suited for  
     reconnection. A simple hypothesis is, then, that \jettwo\ is associated
with reconnection linked to this topological complexity.

\section{Discussion}

The observational characteristics of the present jet, combined
with its coronal magnetic field environment as derived from the
NLFFF extrapolation, pose a real challenge for an 
interpretation  in terms of the classical jet models. 
In the following we discuss separate aspects of the jet that  can 
provide some guidance in relation to the questions posed in the
introduction. 


\subsection{Reconnection at multiple sites that leads to the jet} \label{sec:disc1}

The jet studied in this paper seems to be launched from a variety
of sites in the boundary area between the large-scale closed
active-region loops and the open field outside. The particular
region on which  we  focused  our analysis is the supergranule
highlighted in the rectangle of \fig{fig:aiahmi}. The
photospheric field in this supergranule is predominantly of
negative polarity, but has a number of weaker positive-polarity
intrusions in its interior. The result, as shown in
Sec.~\ref{sec:topology}, is a complicated field line geometry and
topology. We  located (a) several bald patches with field
lines connected with the jet region; (b) a singular point
similar  to an  asymmetric null point that is connected to all
important domains in the active region (and in the supergranule
itself) and (c) two large, highly-inclined twisted flux ropes
located on either side of the lower spine of the  null-point
structure and with opposite helicity sign. We believe that
reconnection associated with these structures is the initial cause
for the jet and its various strands.

\begin{figure*}
\centering
\hbox to \hsize{
  \includegraphics[width=0.51\textwidth]{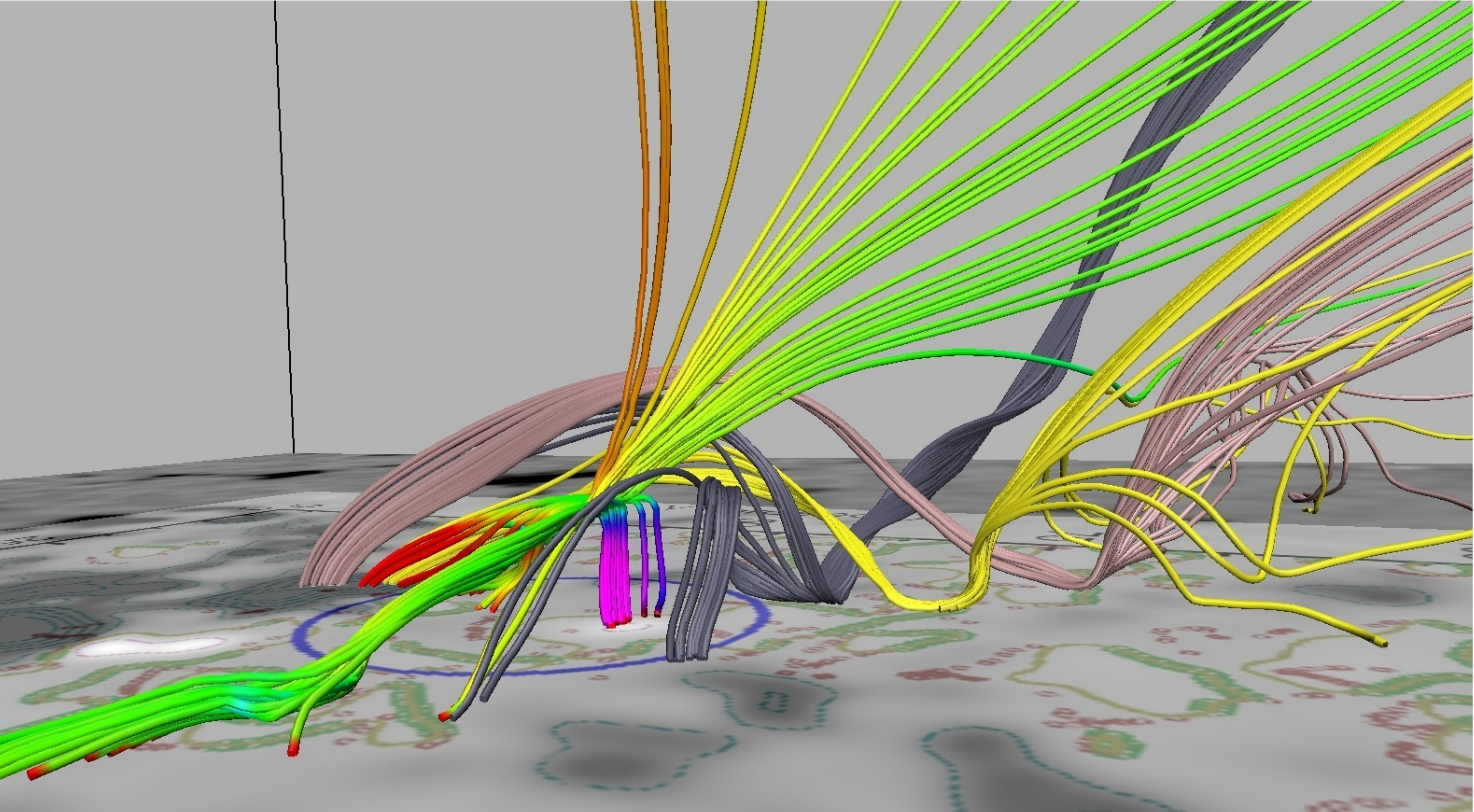}
\includegraphics[width=0.47\textwidth]{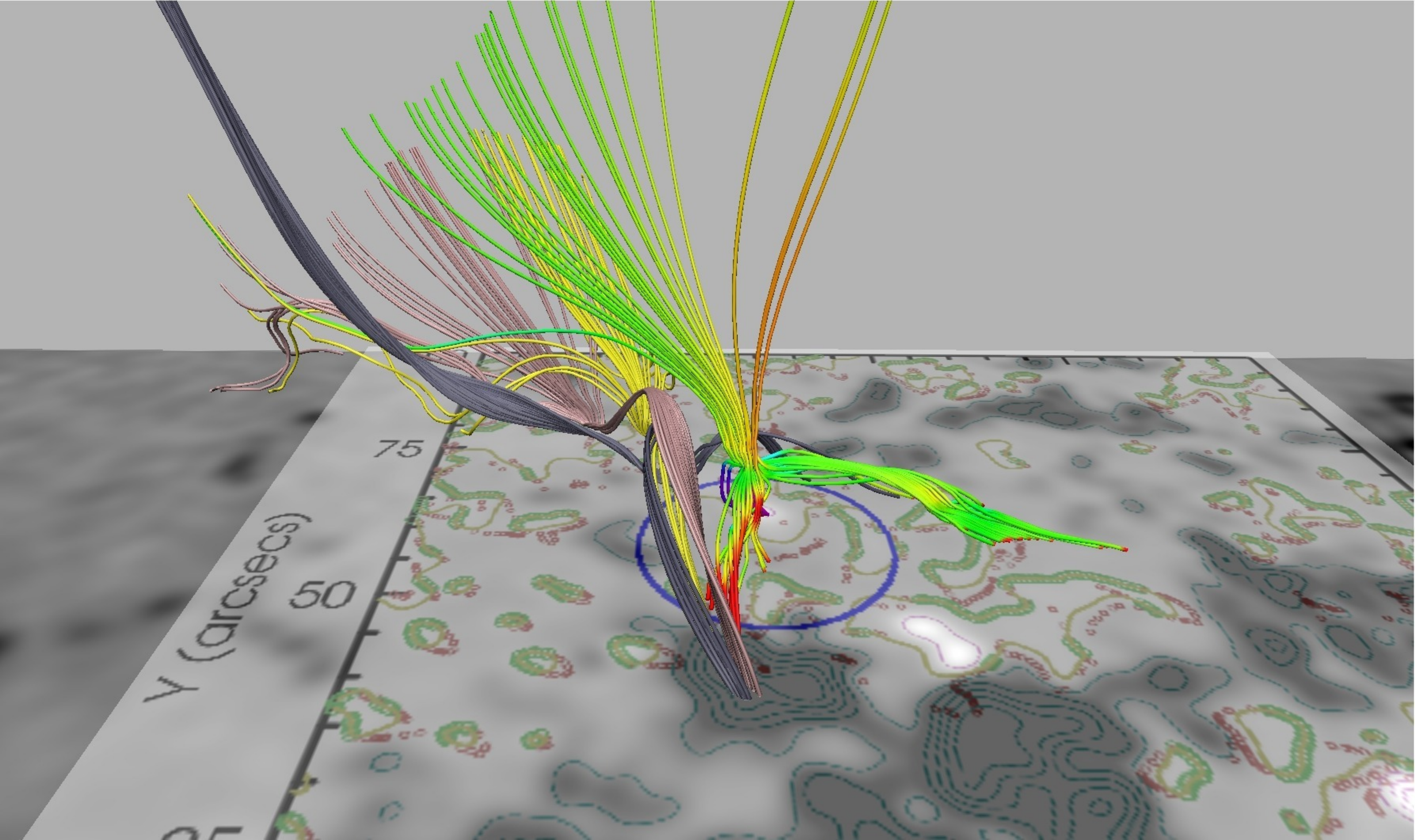}
}
\caption{Visualization of three field line sets that go through the bald
  patches to the north-east of p3 (in dark gray, yellow, and light
  magenta). The two panels show different perspectives of the same field
  lines. The bottom plane contains a superposition of the NLFFF magnetogram
  and the image of \fig{fig:bald}. To facilitate identification, the
  field line set visible in the bottom panels of \fig{fig:vis1} that
 is colored according to the value of $B_z$ is repeated here.}
 \label{fig:bald_patch_and_visualization}
\end{figure*}

The reason why the NLFFF displays several bald patches is that the sparse
photospheric small flux concentrations (p3, p4 and n1, n2) in which the
coronal jet is apparently rooted were actually located in the middle of a
supergranule, where the ambient vertical flux density was weak. Additionally,
a significant part (but not all) of the remote connectivity of the
supergranule interior is  done with the field in the outskirts of the active
region, which has a substantial inclination from the vertical direction here.
This roughly horizontal field was stronger in magnitude than that of the
small flux concentrations like p3 or p4. The latter, then, cause  small
perturbations to this ambient field. The p3 positive polarity was strong
enough to produce a deep minimum of the field strength above itself, which
can be identified with 
a null point with real eigenvalues. No other singular structures of null-point
 type were found in the region, there is an abundance of bald
patches in it. The remaining remote connectivity of the supergranule interior
is with the positive polarity of the active region via tall coronal loops: in
fact one sees in neighboring patches a mixture between the two types of
remote connectivity (to the AR positive polarity or to the field in the
outskirts of the active region, respectively). This is particularly  the
case of the topology near p3, as shown in \fig{fig:vis1}, bottom-left panel.

Therefore, while we cannot pinpoint any individual separatrix layer at a
specific location as directly responsible for the reconnection
that leads to the jet, we argue that the appearance of the weaker
positive polarities in the supergranule interior and the general
horizontal motions in the supergranule, as exemplified in the LCT
map of \fig{fig:lct}, may have generated electric currents at the
separatrices (see \citealt{1988Low}, \citealt{1993Billinghurst},
\citealt{2009Pariat2}), until the jet reconnection eventually
started, in particular related to the p3 positive polarity with
its neighboring bald patches in the north-east and west.  The
fact that the jet has several components that start at different
times and propagate at different velocities (as is clear from
the time-slice analysis along the jet in 171~\AA ) is an indication
that different reconnection sites were sequentially activated
during the event, probably because several of them were linked by
some common field lines. 
On the other hand, the similarity between the field-line geometry  shown
in Figures~\ref{fig:vis1} and \ref{fig:bald_patch_and_visualization} and the
general shape of \jettwo\ apparent in \fig{fig:aiahmi} 
supports the assumption that the field-line sets and structures identified in
Section~\ref{sec:topology} are directly associated with the jet, in spite of
the uncertainty of the  magnetic field inversion results used in this study
(especially in weak field-regions) and the general limitations of NLFF
modeling.

\subsection{Conversion of flux ropes into torsional Alfv\'en waves} \label{sec:disc2}

Different recent observations of X-ray or EUV jets, referred to
in Sect.~1, have revealed the occurrence of (un)twisting motions
at high altitude along the jets \citep[the first of which
 was ][]{2008Patsourakos}.  These have been interpreted as the
propagation of a torsional Alfv\'en wave generated by the
reconnection between large-scale loops and low-lying twisted
fields (firstly by \citealt{2009Pariat}, then confirmed by
\citealt{2009Torok}). Suggestions of MHD waves propagating along
jets have also been provided by \citet{2008Nishizuka}. But thus
far direct evidence of magnetic twist existing before the
observed launch of jets has been elusive; a possible exception is
the work of \citet{2012Sun}, but in their case the twisted
coronal field lines actually erupted all together, as in the
mini-CME eruptions reported for the violent phase of the blowout
jets \citep[as in][]{2010Moore, 2010Sterling}. From the point of
view of theory, the eruption of twisted ropes seems to be a
natural process occurring in the aftermath of a quiescent jet
ejection: twisted ropes created at the base of the jet erupt
through a variety of mechanisms, thus bringing twisted field
lines to  higher levels \citep{2013Moreno-Insertis,
  2013ApJ...769L..21A}, thereby possibly providing an explanation
for the blowout-jet phenomenon.

The jet studied in this paper seems to share common elements with
both the quiescent and violent phases of the blowout-jet
phenomenon. The broad nature of the jet and its transient
undulatory transverse motions (as shown in the 171~\AA
\ time-slice plots across the jet of Sec.~\ref{sec:slice}) are
suggestive of the propagation of (possibly torsional) Alfv\'en
waves. The NLFFF extrapolation offers a natural explanation for
the generation of these waves. The two highly inclined flux ropes
on either side of the positive flux patch p3 discussed in
Sec.~\ref{sec:topology} must have been created in relation to the
formation of that positive polarity itself (as corresponds to the
opposite signs of the helicity in these ropes).  The field lines
in the ropes are actually rooted in different regions of the
supergranule interior; in the other direction, they are part of
the outward-going field lines at the AR periphery.  The
reconnection creating the ropes then must lead to a twist
imbalance between the ropes' low- and high-altitude sections. This
will subsequently lead to the upward propagation of a torsional
Alfv\'en wave. Depending on the circumstances, the ropes
themselves may erupt, or, alternatively, all or most of their
pre-eruptive twist may be carried away with the torsional wave
and the ropes do not erupt.  This means that this jet may have features
corresponding to the two phases of the blowout-jet phenomenon.

Observationally, however, there is no definitive evidence for any
of the possibilities just mentioned. On the positive side, the
pre-event corona around the footpoints of the jet does show
complex 171~\AA \ emission patterns varying on a much smaller
spatial scale than where large coronal loops can be seen. But
these patterns cannot be indisputably attributed to the presence
of a small-scale low-lying flux rope. In addition, we do not see
a clear flux-rope-eruption-like signature in the event.

We therefore   conjecture that the strength and orientation of
the ambient magnetic field in a region where a jet is launched
following the appearance of new magnetic polarities has a strong effect on the nature and on the observational
characteristics of the jet (as previously shown for a simple
configuration by \citealt{2007Galsgaard}). For instance, if the
contrast of orientations between ambient field and emerging flux
is high enough and the latter has sufficient field strength and
total flux,  a coronal null point may result and the jet may
be of the type described by the three dimensional  numerical models mentioned in
the introduction. On the other hand, if the orientation contrast
is not too strong and/or the emerging field is too weak, a bald
patch may result and the reconnection may be mostly associated
with the latter structure, possibly in the lower part of the
corona.  Also, if a strong-enough horizontal flux rope with
sufficient twist is initially present, one can obtain an eruption
as in the violent phase of the blowout jets. But if the flux rope
is too weak,  it will be converted into a torsional wave by
reconnection, which will thus result in a classical untwisting
jet.

The foregoing results raise a number of questions concerning the
  generality of the jet-launching mechanisms that lie at the basis of the
  event analyzed in this paper. From the theoretical point of view, the
  necessity and mutual influence of flux emergence, organization of the field
  in bald-patch regions, creation of the flux ropes, etc. in causing the jets
  can only be understood from a time-evolution analysis of a complex
  region. Observationally, one can only ascertain the generality of this type
  of jets as a separate class via a statistical analysis of coronal and
  photospheric data. These are natural extensions of this work that go beyond
  the objectives of the current paper.


\begin{acknowledgements}
We thank E. Pariat, R. Centeno, M. Cheung, P. D\'emoulin, and
  K. Galsgaard for fruitful discussions. This work has been initiated at the
ISSI (Bern) during the workshop {\it Magnetic flux emergence in the solar
  atmosphere} organized by K.Galsgaard and F. Zucarello. ISSI support for
that workshop and for the workshop {\it Understanding Solar Jets and their
  Role in Atmospheric Structure and Dynamics} organized by N.E. Raouafi and
E. Pariat is acknowledged.  YG was supported by the National Natural Science
Foundation of China (NSFC) under the grant numbers 11203014, 10933003, and
the grant from the 973 project 2011CB811402. FMI and LYC have been supported
through grants AYA2011-24808 and CSD2007-00050 of the Spanish Ministry of
Economy. We gratefully acknowledge the computer resources and
  assistance provided at the MareNostrum (BSC/CNS/RES, Spain) and LaPalma
  (IAC/RES, Spain). We are also grateful for the use of UCAR's
  VAPOR 3D visualization package
  \citep{2007NJPh....9..301C}. JKT acknowledges support from DFG grant WI 
3211/2-1. YG thanks the Observatoire de Paris for the grant it was given during his
stay in Meudon in February 2013. SDO AIA and HMI data are courtesy of the
NASA/SDO science teams.
\end{acknowledgements}

\bibliographystyle{aa}
\bibliography{bibliography_sci}

\begin{thebibliography}{64}
\expandafter\ifx\csname natexlab\endcsname\relax\def\natexlab#1{#1}\fi

\bibitem[{{Aly}(1989)}]{1989Aly}
{Aly}, J.~J. 1989, \solphys, 120, 19

\bibitem[{{Archontis} \& {Hood}(2012)}]{2012Archontis}
{Archontis}, V. \& {Hood}, A.~W. 2012, \aap, 537, A62

\bibitem[{{Archontis} \& {Hood}(2013)}]{2013ApJ...769L..21A}
{Archontis}, V. \& {Hood}, A.~W. 2013, \apjl, 769, L21

\bibitem[{{Aulanier} {et~al.}(1998){Aulanier}, {D{\'e}moulin}, {Schmieder},
  {Fang}, \& {Tang}}]{1998Aulanier}
{Aulanier}, G., {D{\'e}moulin}, P., {Schmieder}, B., {Fang}, C., \& {Tang},
  Y.~H. 1998, \solphys, 183, 369

\bibitem[{{Billinghurst} {et~al.}(1993){Billinghurst}, {Craig}, \&
  {Sneyd}}]{1993Billinghurst}
{Billinghurst}, M.~N., {Craig}, I.~J.~D., \& {Sneyd}, A.~D. 1993, \aap, 279,
  589

\bibitem[{{Borrero} {et~al.}(2011){Borrero}, {Tomczyk}, {Kubo},
  {Socas-Navarro}, {Schou}, {Couvidat}, \& {Bogart}}]{2011Borrero}
{Borrero}, J.~M., {Tomczyk}, S., {Kubo}, M., {et~al.} 2011, \solphys, 273, 267

\bibitem[{{Bungey} {et~al.}(1996){Bungey}, {Titov}, \& {Priest}}]{1996Bungey}
{Bungey}, T.~N., {Titov}, V.~S., \& {Priest}, E.~R. 1996, \aap, 308, 233

\bibitem[{{Canfield} {et~al.}(1996){Canfield}, {Reardon}, {Leka}, {Shibata},
  {Yokoyama}, \& {Shimojo}}]{1996Canfield}
{Canfield}, R.~C., {Reardon}, K.~P., {Leka}, K.~D., {et~al.} 1996, \apj, 464,
  1016

\bibitem[{{Chen} {et~al.}(2012){Chen}, {Zhang}, \& {Ma}}]{2012Chen}
{Chen}, H.-D., {Zhang}, J., \& {Ma}, S.-L. 2012, Research in Astronomy and
  Astrophysics, 12, 573

\bibitem[{{Cirtain} {et~al.}(2007){Cirtain}, {Golub}, {Lundquist}, {van
  Ballegooijen}, {Savcheva}, {Shimojo}, {DeLuca}, {Tsuneta}, {Sakao}, {Reeves},
  {Weber}, {Kano}, {Narukage}, \& {Shibasaki}}]{2007Cirtain}
{Cirtain}, J.~W., {Golub}, L., {Lundquist}, L., {et~al.} 2007, Science, 318,
  1580

\bibitem[{{Clyne} {et~al.}(2007){Clyne}, {Mininni}, {Norton}, \&
  {Rast}}]{2007NJPh....9..301C}
{Clyne}, J., {Mininni}, P., {Norton}, A., \& {Rast}, M. 2007, New Journal of
  Physics, 9, 301

\bibitem[{{D\'emoulin} {et~al.}(1996){D\'emoulin}, {H\'enoux}, {Priest}, \&
  {Mandrini}}]{1996Demoulin}
{D\'emoulin}, P., {H\'enoux}, J.~C., {Priest}, E.~R., \& {Mandrini}, C.~H.
  1996, \aap, 308, 643

\bibitem[{{DeRosa} {et~al.}(2009){DeRosa}, {Schrijver}, {Barnes}, {Leka},
  {Lites}, {Aschwanden}, {Amari}, {Canou}, {McTiernan}, {R{\'e}gnier},
  {Thalmann}, {Valori}, {Wheatland}, {Wiegelmann}, {Cheung}, {Conlon},
  {Fuhrmann}, {Inhester}, \& {Tadesse}}]{2009DeRosa}
{DeRosa}, M.~L., {Schrijver}, C.~J., {Barnes}, G., {et~al.} 2009, \apj, 696,
  1780

\bibitem[{{Galsgaard} {et~al.}(2007){Galsgaard}, {Archontis},
  {Moreno-Insertis}, \& {Hood}}]{2007Galsgaard}
{Galsgaard}, K., {Archontis}, V., {Moreno-Insertis}, F., \& {Hood}, A.~W. 2007,
  \apj, 666, 516

\bibitem[{{Gary} \& {Hagyard}(1990)}]{1990Gary}
{Gary}, G.~A. \& {Hagyard}, M.~J. 1990, \solphys, 126, 21

\bibitem[{{Guo} {et~al.}(2013){Guo}, {D\'emoulin}, {Schmieder}, {Ding}, {Vargas
  Dom\'inguez}, \& {Liu}}]{2013Guo}
{Guo}, Y., {D\'emoulin}, P., {Schmieder}, B., {et~al.} 2013, A\&A, 55, A19

\bibitem[{{Heyvaerts} {et~al.}(1977){Heyvaerts}, {Priest}, \&
  {Rust}}]{1977Heyvaerts}
{Heyvaerts}, J., {Priest}, E.~R., \& {Rust}, D.~M. 1977, \apj, 216, 123

\bibitem[{{Leka} {et~al.}(2009){Leka}, {Barnes}, {Crouch}, {Metcalf}, {Gary},
  {Jing}, \& {Liu}}]{2009Leka}
{Leka}, K.~D., {Barnes}, G., {Crouch}, A.~D., {et~al.} 2009, \solphys, 260, 83

\bibitem[{{Lemen} {et~al.}(2012){Lemen}, {Title}, {Akin}, {Boerner}, {Chou},
  {Drake}, {et~al.}}]{2012Lemen}
{Lemen}, J.~R., {Title}, A.~M., {Akin}, D.~J., {et~al.} 2012, \solphys, 275, 17

\bibitem[{{Liu} {et~al.}(2011){Liu}, {Deng}, {Liu}, {Ugarte-Urra}, {Wang}, \&
  {Wang}}]{2011Liu}
{Liu}, C., {Deng}, N., {Liu}, R., {et~al.} 2011, \apjl, 735, L18

\bibitem[{{Low} \& {Wolfson}(1988)}]{1988Low}
{Low}, B.~C. \& {Wolfson}, R. 1988, \apj, 324, 574

\bibitem[{{Madjarska}(2011)}]{2011A&A...526A..19M}
{Madjarska}, M.~S. 2011, \aap, 526, A19

\bibitem[{{Manchester} {et~al.}(2004){Manchester}, {Gombosi}, {DeZeeuw}, \&
  {Fan}}]{2004ApJ...610..588M}
{Manchester}, IV, W., {Gombosi}, T., {DeZeeuw}, D., \& {Fan}, Y. 2004, \apj,
  610, 588

\bibitem[{{Mandrini} {et~al.}(2002){Mandrini}, {D{\'e}moulin}, {Schmieder},
  {Deng}, \& {Rudawy}}]{2002Mandrini}
{Mandrini}, C.~H., {D{\'e}moulin}, P., {Schmieder}, B., {Deng}, Y.~Y., \&
  {Rudawy}, P. 2002, \aap, 391, 317

\bibitem[{{Mandrini} {et~al.}(1996){Mandrini}, {D{\'e}moulin}, {van
  Driel-Gesztelyi}, {Schmieder}, {Cauzzi}, \& {Hofmann}}]{1996Mandrini}
{Mandrini}, C.~H., {D{\'e}moulin}, P., {van Driel-Gesztelyi}, L., {et~al.}
  1996, \solphys, 168, 115

\bibitem[{{McIntosh} \& {De Pontieu}(2009)}]{2009McIntosh}
{McIntosh}, S.~W. \& {De Pontieu}, B. 2009, \apjl, 706, L80

\bibitem[{{Metcalf}(1994)}]{1994Metcalf}
{Metcalf}, T.~R. 1994, \solphys, 155, 235

\bibitem[{{Metcalf} {et~al.}(1995){Metcalf}, {Jiao}, {McClymont}, {Canfield},
  \& {Uitenbroek}}]{1995Metcalf}
{Metcalf}, T.~R., {Jiao}, L., {McClymont}, A.~N., {Canfield}, R.~C., \&
  {Uitenbroek}, H. 1995, \apj, 439, 474

\bibitem[{{Metcalf} {et~al.}(2006){Metcalf}, {Leka}, {Barnes}, {Lites},
  {Georgoulis}, {Pevtsov}, {et~al.}}]{2006Metcalf}
{Metcalf}, T.~R., {Leka}, K.~D., {Barnes}, G., {et~al.} 2006, \solphys, 237,
  267

\bibitem[{{Moore} {et~al.}(2010){Moore}, {Cirtain}, {Sterling}, \&
  {Falconer}}]{2010Moore}
{Moore}, R.~L., {Cirtain}, J.~W., {Sterling}, A.~C., \& {Falconer}, D.~A. 2010,
  \apj, 720, 757

\bibitem[{Moreno-Insertis \& Galsgaard(2013)}]{2013Moreno-Insertis}
Moreno-Insertis, F. \& Galsgaard, K. 2013, The Astrophysical Journal, 771, 20

\bibitem[{{Moreno-Insertis} {et~al.}(2008){Moreno-Insertis}, {Galsgaard}, \&
  {Ugarte-Urra}}]{2008Moreno-Insertis}
{Moreno-Insertis}, F., {Galsgaard}, K., \& {Ugarte-Urra}, I. 2008, \apjl, 673,
  L211

\bibitem[{{Nishizuka} {et~al.}(2008){Nishizuka}, {Shimizu}, {Nakamura},
  {Otsuji}, {Okamoto}, {Katsukawa}, \& {Shibata}}]{2008Nishizuka}
{Nishizuka}, N., {Shimizu}, M., {Nakamura}, T., {et~al.} 2008, \apjl, 683, L83

\bibitem[{{November} \& {Simon}(1988)}]{1988November}
{November}, L.~J. \& {Simon}, G.~W. 1988, \apj, 333, 427

\bibitem[{{Pariat} {et~al.}(2009{\natexlab{a}}){Pariat}, {Antiochos}, \&
  {DeVore}}]{2009Pariat}
{Pariat}, E., {Antiochos}, S.~K., \& {DeVore}, C.~R. 2009{\natexlab{a}}, \apj,
  691, 61

\bibitem[{{Pariat} {et~al.}(2010){Pariat}, {Antiochos}, \&
  {DeVore}}]{2010Pariat}
{Pariat}, E., {Antiochos}, S.~K., \& {DeVore}, C.~R. 2010, \apj, 714, 1762

\bibitem[{{Pariat} {et~al.}(2004){Pariat}, {Aulanier}, {Schmieder},
  {Georgoulis}, {Rust}, \& {Bernasconi}}]{2004Pariat}
{Pariat}, E., {Aulanier}, G., {Schmieder}, B., {et~al.} 2004, \apj, 614, 1099

\bibitem[{{Pariat} {et~al.}(2009{\natexlab{b}}){Pariat}, {Masson}, \&
  {Aulanier}}]{2009Pariat2}
{Pariat}, E., {Masson}, S., \& {Aulanier}, G. 2009{\natexlab{b}}, \apj, 701,
  1911

\bibitem[{{Patsourakos} {et~al.}(2008){Patsourakos}, {Pariat}, {Vourlidas},
  {Antiochos}, \& {Wuelser}}]{2008Patsourakos}
{Patsourakos}, S., {Pariat}, E., {Vourlidas}, A., {Antiochos}, S.~K., \&
  {Wuelser}, J.~P. 2008, \apjl, 680, L73

\bibitem[{{Pontin}(2011)}]{2011AdSpR..47.1508P}
{Pontin}, D.~I. 2011, Advances in Space Research, 47, 1508

\bibitem[{{Savcheva} {et~al.}(2007){Savcheva}, {Cirtain}, {Deluca},
  {Lundquist}, {Golub}, {Weber}, {Shimojo}, {Shibasaki}, {Sakao}, {Narukage},
  {Tsuneta}, \& {Kano}}]{2007Savcheva}
{Savcheva}, A., {Cirtain}, J., {Deluca}, E.~E., {et~al.} 2007, \pasj, 59, 771

\bibitem[{{Scherrer} {et~al.}(2012){Scherrer}, {Schou}, {Bush}, {Kosovichev},
  {Bogart}, {Hoeksema}, {et~al.}}]{2012Scherrer}
{Scherrer}, P.~H., {Schou}, J., {Bush}, R.~I., {et~al.} 2012, \solphys, 275,
  207

\bibitem[{{Schmieder} {et~al.}(1997){Schmieder}, {Aulanier}, {Demoulin}, {van
  Driel-Gesztelyi}, {Roudier}, {Nitta}, \& {Cauzzi}}]{1997Schmieder}
{Schmieder}, B., {Aulanier}, G., {Demoulin}, P., {et~al.} 1997, \aap, 325, 1213

\bibitem[{{Schmieder} {et~al.}(1995){Schmieder}, {Shibata}, {van
  Driel-Gesztelyi}, \& {Freeland}}]{1995Schmieder}
{Schmieder}, B., {Shibata}, K., {van Driel-Gesztelyi}, L., \& {Freeland}, S.
  1995, \solphys, 156, 245

\bibitem[{{Schou} {et~al.}(2012){Schou}, {Scherrer}, {Bush}, {Wachter},
  {Couvidat}, {Rabello-Soares}, {et~al.}}]{2012Schou}
{Schou}, J., {Scherrer}, P.~H., {Bush}, R.~I., {et~al.} 2012, \solphys, 275,
  229

\bibitem[{{Shen} {et~al.}(2011){Shen}, {Liu}, {Su}, \& {Ibrahim}}]{2011Shen}
{Shen}, Y., {Liu}, Y., {Su}, J., \& {Ibrahim}, A. 2011, \apjl, 735, L43

\bibitem[{{Shibata} {et~al.}(1992){Shibata}, {Nozawa}, \&
  {Matsumoto}}]{1992Shibata265}
{Shibata}, K., {Nozawa}, S., \& {Matsumoto}, R. 1992, \pasj, 44, 265

\bibitem[{{Shimojo} {et~al.}(1996){Shimojo}, {Hashimoto}, {Shibata},
  {Hirayama}, {Hudson}, \& {Acton}}]{1996Shimojo}
{Shimojo}, M., {Hashimoto}, S., {Shibata}, K., {et~al.} 1996, \pasj, 48, 123

\bibitem[{{Shimojo} \& {Shibata}(2000)}]{2000Shimojo}
{Shimojo}, M. \& {Shibata}, K. 2000, \apj, 542, 1100

\bibitem[{{Shimojo} {et~al.}(1998){Shimojo}, {Shibata}, \&
  {Harvey}}]{1998Shimojo}
{Shimojo}, M., {Shibata}, K., \& {Harvey}, K.~L. 1998, \solphys, 178, 379

\bibitem[{{Sterling} {et~al.}(2010){Sterling}, {Harra}, \&
  {Moore}}]{2010Sterling}
{Sterling}, A.~C., {Harra}, L.~K., \& {Moore}, R.~L. 2010, \apj, 722, 1644

\bibitem[{{Sun} {et~al.}(2012){Sun}, {Hoeksema}, {Liu}, {Chen}, \&
  {Hayashi}}]{2012Sun}
{Sun}, X., {Hoeksema}, J.~T., {Liu}, Y., {Chen}, Q., \& {Hayashi}, K. 2012,
  \apj, 757, 149

\bibitem[{{Tiwari}(2012)}]{2012Tiwari}
{Tiwari}, S.~K. 2012, \apj, 744, 65

\bibitem[{{T{\"o}r{\"o}k} {et~al.}(2009){T{\"o}r{\"o}k}, {Aulanier},
  {Schmieder}, {Reeves}, \& {Golub}}]{2009Torok}
{T{\"o}r{\"o}k}, T., {Aulanier}, G., {Schmieder}, B., {Reeves}, K.~K., \&
  {Golub}, L. 2009, \apj, 704, 485

\bibitem[{{T{\"o}r{\"o}k} \& {Kliem}(2003)}]{2003A&A...406.1043T}
{T{\"o}r{\"o}k}, T. \& {Kliem}, B. 2003, \aap, 406, 1043

\bibitem[{{Tsuneta} {et~al.}(1991){Tsuneta}, {Acton}, {Bruner}, {Lemen},
  {Brown}, {Caravalho}, {Catura}, {Freeland}, {Jurcevich}, {Morrison},
  {Ogawara}, {Hirayama}, \& {Owens}}]{1991Tsuneta}
{Tsuneta}, S., {Acton}, L., {Bruner}, M., {et~al.} 1991, \solphys, 136, 37

\bibitem[{{Ugarte-Urra} \& {Warren}(2011)}]{2011Ugarte-Urra}
{Ugarte-Urra}, I. \& {Warren}, H.~P. 2011, \apj, 730, 37

\bibitem[{{Wheatland} {et~al.}(2000){Wheatland}, {Sturrock}, \&
  {Roumeliotis}}]{2000Wheatland}
{Wheatland}, M.~S., {Sturrock}, P.~A., \& {Roumeliotis}, G. 2000, \apj, 540,
  1150

\bibitem[{{Wiegelmann}(2004)}]{2004Wiegelmann}
{Wiegelmann}, T. 2004, \solphys, 219, 87

\bibitem[{{Wiegelmann} \& {Inhester}(2010)}]{2010Wiegelmann}
{Wiegelmann}, T. \& {Inhester}, B. 2010, \aap, 516, A107

\bibitem[{{Wiegelmann} {et~al.}(2006){Wiegelmann}, {Inhester}, \&
  {Sakurai}}]{2006Wiegelmann}
{Wiegelmann}, T., {Inhester}, B., \& {Sakurai}, T. 2006, \solphys, 233, 215

\bibitem[{{Yokoyama} \& {Shibata}(1995)}]{1995Yokoyama}
{Yokoyama}, T. \& {Shibata}, K. 1995, \nat, 375, 42

\bibitem[{{Yokoyama} \& {Shibata}(1996)}]{1996Yokoyama}
{Yokoyama}, T. \& {Shibata}, K. 1996, \pasj, 48, 353

\bibitem[{{Zhang} {et~al.}(2012){Zhang}, {Chen}, {Guo}, {Fang}, \&
  {Ding}}]{2012Zhang}
{Zhang}, Q.~M., {Chen}, P.~F., {Guo}, Y., {Fang}, C., \& {Ding}, M.~D. 2012,
  \apj, 746, 19

\end{thebibliography}

\end{document}